\documentclass{aastex62}

\shorttitle{\WFIRST Prism}
\shortauthors{Rubin et al.}

\usepackage[utf8]{inputenc}
\usepackage{xspace}
\usepackage{hyperref}
\usepackage[acronym, nopostdot]{glossaries}
\setacronymstyle{long-short}
\usepackage{enumerate}

\makeglossaries

\begin{document}

\newacronym{SED}{SED}{spectral energy distribution}
\newacronym{SALT2}{SALT2}{Spectral Adaptive Light-curve Template}
\newacronym{SUGAR}{SUGAR}{SUpernova Generator And Reconstructor}
\newacronym{SNEMO}{SNEMO}{SuperNova Empirical MOdels}
\newacronym{SNfactory}{SNfactory}{Nearby Supernova Factory}
\newacronym{LSF}{LSF}{Line-Spread Function}
\newacronym{PSF}{PSF}{Point-Spread Function}
\newacronym{WFI}{WFI}{Wide Field Instrument}
\newacronym{JWST}{{\it JWST}}{the {\it James Webb Space Telescope}}
\newacronym{SDT}{SDT}{Science Definition Team}
\newacronym{PFS}{PFS}{Prime Focus Spectrograph}
\newacronym{FoM}{FoM}{Dark Energy Task Force Figure of Merit}
\newacronym{ParSNIP}{ParSNIP}{Parameterization of SuperNova Intrinsic Properties}
\newacronym{SNID}{SNID}{Supernova Identification}
\newacronym{CMB}{CMB}{Cosmic Microwave Background}
\newacronym{HLTDS}{HLTDS}{High Latitude Time Domain Survey}
\newacronym{EMPCA}{EM-PCA}{Expectation-Maximization Principal Component Analysis}

\newcommand{\Ang}{\AA\xspace}
\newcommand{\WFIRST}{{\it Roman}\xspace}
\newcommand{\WFIRSTspelled}{{\it Nancy Grace Roman Space Telescope}\xspace}

\title{\large{Evaluating and Optimizing a Slitless Prism for \WFIRSTspelled SN Cosmology}}

\newcommand{\uhawaii}{\affiliation{Department of Physics and Astronomy, University of Hawai`i at M{\=a}noa, Honolulu, Hawai`i 96822}}
\newcommand{\stsci}{\affiliation{Space Telescope Science Institute, 3700 San Martin Drive Baltimore, MD 21218, USA}}
\newcommand{\lbnl}{\affiliation{E.O. Lawrence Berkeley National Laboratory, 1 Cyclotron Rd., Berkeley, CA, 94720, USA}}
\newcommand{\yale}{\affiliation{Department of Physics, Yale University, New Haven, CT, 06250-8121, USA}}
\newcommand{\ucb}{\affiliation{Department of Physics, University of California Berkeley, 366 LeConte Hall MC 7300, Berkeley, CA, 94720, USA}}
\newcommand{\duke}{\affiliation{Department of Physics, Duke University, 120 Science Drive, Durham, NC, 27708, USA}}
\newcommand{\gsfc}{\affiliation{NASA Goddard Space Flight Center, Greenbelt, MD 20771, USA}}
\newcommand{\umbc}{\affiliation{University of Maryland, Baltimore County, Baltimore, MD 21250, USA}}

\author[0000-0001-5402-4647]{David Rubin}
\uhawaii
\stsci
\lbnl

\author{Greg Aldering}
\lbnl

\author{Tri L. Astraatmadja}
\stsci

\author{Charlie Baltay}
\yale

\author[0000-0001-7101-9831]{Aleksandar Cikota}
\affiliation{European Southern Observatory, Alonso de C\'{o}rdova 3107, Casilla
19, Santiago, Chile}
\lbnl

\author[0000-0003-2823-360X]{Susana E.\ Deustua}
\affiliation{Sensor Science Division, National Institute of Standards and Technology, 100 Bureau Dr., Gaithersburg, MD 20899-8444, USA}

\author{Sam Dixon}
\lbnl
\ucb

\author{Andrew Fruchter}
\stsci

\author[0000-0002-1296-6887]{L. Galbany}
\affiliation{Institute of Space Sciences (ICE, CSIC), Campus UAB, Carrer de Can Magrans, s/n, E-08193 Barcelona, Spain.}
\affiliation{Institut d’Estudis Espacials de Catalunya (IEEC), E-08034 Barcelona, Spain.}

\author{Rebekah Hounsell}
\umbc
\gsfc

\author{Saul Perlmutter}
\lbnl
\ucb

\author{Ben Rose}
\duke

\correspondingauthor{David Rubin}
\email{drubin@hawaii.edu}

\newcommand{\LCDM}{\ensuremath{\Lambda}CDM\xspace}
\newcommand{\note}[1]{{\textcolor{red}{#1}}\xspace}

\newcommand{\firstsentence}{This work presents a set of studies addressing the use of the low-dispersion slitless prism on \WFIRST for SN spectroscopy as part of the \WFIRST \gls{HLTDS}.\xspace}
\newcommand{\prismsurveyoptimalsummary}{for surveys taking 0.125 years of prism time, the optimum is a $\sim 5$~deg$^2$ wide tier with $\sim 600$~s pointings, with a $\sim 1$~deg$^2$ deep tier with $\sim 1$~hour pointings\xspace}

\newcommand{\NumTiers}{two\xspace}
\newcommand{\TierOne}{wide\xspace}
\newcommand{\TierTwo}{deep\xspace}
\newcommand{\TierSlash}{\TierOne/\TierTwo}

\newcommand{\numberpoints}{three\xspace}

\newcommand{\widedegsq}{5.32\xspace}
\newcommand{\widepointings}{19~pointings\xspace}
\newcommand{\deepdegsq}{1.12\xspace}
\newcommand{\deeppointings}{4~pointings\xspace}
\newcommand{\SqDegSlash}{\widedegsq square degree and \deepdegsq square degree\xspace}

\newcommand{\widehours}{3.5\xspace} 
\newcommand{\deephours}{4\xspace} 
\newcommand{\wideexp}{600\xspace}
\newcommand{\deepexp}{3600\xspace}
\newcommand{\slewtime}{62.15 seconds (22 detector readouts of 2.825~seconds each)\xspace}
\newcommand{\imagingonlyseconds}{2410.75~s\xspace}
\newcommand{\prismonlyseconds}{2348.6~s\xspace}

\newcommand{\waverange}{0.5 to 0.6 $\mu$m\xspace}
\newcommand{\obswaverange}{0.75 to 1.8 $\mu$m\xspace}
\newcommand{\gooddispersion}{$\gtrsim 70$\xspace}

\newcommand{\multiplexword}{$\sim 10$\xspace}
\newcommand{\redshiftsubtype}{$z \sim 1.3$\xspace}

\newcommand{\NSNefifteen}{2500\xspace}
\newcommand{\NSNefifteenHost}{1800\xspace}

\newcommand{\NSNefifteentotwentyfive}{1000\xspace} 

\newcommand{\NSNetwentyfive}{1500\xspace}
\newcommand{\NSNetwentyfiveHost}{940\xspace}
\newcommand{\NSNethirtyfive}{960\xspace}
\newcommand{\NSNethirtyfiveHost}{530\xspace}
\newcommand{\NSNefifty}{520\xspace}
\newcommand{\NSNefiftyHost}{240\xspace}

\newcommand{\asf}[1]{\textcolor{purple}{#1}} 
\newcommand{\alg}[1]{\textcolor{blue}{#1}} 
\newcommand{\jan}[1]{\textcolor{purple}{#1}} 
\newcommand{\bmr}[1]{\textcolor{red}{#1}}

\submitjournal{ApJ}

\begin{abstract}

\firstsentence We find SN spectral energy distributions including prism data carry more information than imaging alone at fixed total observing time, improving redshift measurements and sub-typing of SNe. The \WFIRST field of view will typically include \multiplexword SNe~Ia at observable redshifts at a range of phases (the multiplexing of host galaxies is much greater as they are always present), building up SN spectral time series without targeted observations. We show that fitting these time series extracts more information than stacking the data over all the phases, resulting in a large improvement in precision for SN~Ia subclassification measurements. A prism on \WFIRST thus significantly enhances scientific opportunities for the mission, and is particularly important for the \WFIRST SN cosmology program to provide the systematics-controlled measurement that is a focus of the \WFIRST dark energy mission. Optimizing the prism parameters, we conclude that the blue cutoff should be set as blue as the prism image quality allows ($\sim 7500$\Ang), the red cutoff should be set to $\sim 18000$\Ang to minimize thermal background, and the two-pixel dispersion should be \gooddispersion.
\end{abstract}

\glsresetall

\section{Introduction} \label{sec:introduction}

\begin{figure}[htbp]
    \centering
    \includegraphics[width=\textwidth]{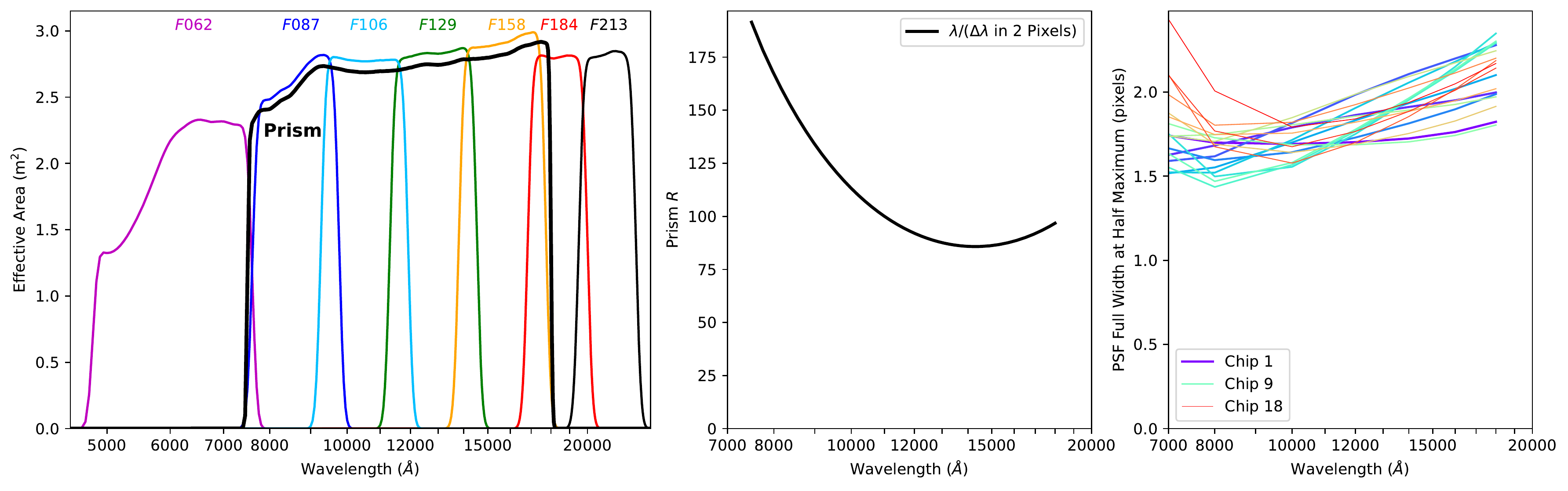}
    \caption{{\bf Left panel}: the effective area of the \gls{WFI} moderate-width imaging filters and the prism (not shown: the wide $F146$ and the grism). {\bf Center panel}: two-pixel dispersion $R$ for the prism. {\bf Right panel}: prism \gls{PSF} full width at half maximum in pixels as a function of wavelength for each of the 18 detectors (evaluated with a 2D Gaussian fit after convolution with the pixel). Generally speaking, the \gls{PSF} is reasonably well-sampled and the two-pixel dispersion is a reasonably accurate representation of the resolution.}
    \label{fig:filters}
\end{figure}

The \WFIRSTspelled (\WFIRST) is an under-construction flagship space telescope designed for coronagraphy and wide-field optical-NIR observations \citep{spergel15}. The \gls{WFI} is the baseline imaging and slitless spectroscopic instrument for \WFIRST. The \gls{WFI} will observe 0.28 square degrees per pointing with $0\farcs 11$ pixels in seven moderate-width filters (shown in Figure~\ref{fig:filters}) as well as one very wide filter and grism.

One of the primary goals of \WFIRST is to investigate the dynamics of the accelerated expansion of the universe, reaching a 10$\times$ improvement in \gls{FoM} compared to today.\footnote{For example, SNe Ia with external \gls{CMB} measurements should reach a \gls{FoM} of 325 with all uncertainties included (Science Requirement SN 2.0.1).}  Such measurements have the potential to revolutionize our understanding of cosmology. To achieve these ambitious science objectives, \WFIRST will employ several cosmological probes, including measuring luminosity distances of Type~Ia supernovae (SNe Ia), the technique being studied by the two SN-focused Science Investigation Teams.

In current measurements, the known statistical and systematic uncertainties are generally comparably sized \citep{Brout2022}, so only increasing the number of SNe will not yield the desired small cosmological uncertainties. Furthermore, there already exist modest tensions between techniques \citep[e.g.,][]{delubac15, hikage18, PlanckCollaboration2020, Riess2021}, either pointing to the need for even more complicated cosmological models or the presence of undetected systematic errors \citep[e.g.,][]{DiValentino2021}. Thus, control of systematic uncertainties will be crucial for \WFIRST cosmology.

\firstsentence The value of spectroscopy for observing live SNe comes down to \numberpoints major points: redshift measurements, spectroscopic SN classifications, and making spectrophotometric measurements of the SN population.\footnote{This work does not consider two other possible uses of the prism for the \gls{HLTDS}: 1) obtaining redshifts of SN host galaxies and 2) observing brighter spectrophotometric standard stars than are possible to observe with imaging, thus improving the absolute calibration by tying it to more or better-understood standard stars. 
} Most high-redshift SN Ia surveys are not able to spectroscopically observe all of their SNe with good light curves (e.g., \citealt{Smith2020}). If a similar survey design is adopted for \WFIRST, (i.e., a prism survey area smaller than the imaging survey area), then each of these points takes on a dual role: applying to both the direct subset of SNe observed with the prism, and to provide more detailed dataset for training and validating analyses of SNe only observed with the imaging. We elaborate on these three points below:

\begin{enumerate}[I]
\item Redshift measurements of live SNe to allow accurate placement of SNe on the Hubble diagram. For the training role of the prism, these measurements help assess performance and biases of both photometric redshifts \citep{Roberts2017} and association of transients with nearby galaxies that may be the host \citep{Gupta2016}.

\item Spectroscopic classification of transients. These classifications enable an initial spectroscopic-only cosmology analysis (e.g., \citealt{Abbott2019} for the Dark Energy Survey). They also can validate and train photometric classifiers with a random sample of high-redshift events.

\item Finally, and perhaps most importantly, providing spectrophotometric constraints on the \gls{SED} distribution of SNe at high redshifts. Again, this enables an initial cosmology analysis based on subclassifying SNe~Ia for better distance precision (e.g., \citealt{twins15, Boone2021b}). This sample also provides a high-redshift check on population evolution and other systematic uncertainties in the larger photometric-only sample \citep[e.g.,][]{Sullivan2009}. As we show, prism data can even train any components of \gls{SED} variation that may not be present in current low-redshift spectrophotometric training sets.
\end{enumerate}

Certainly by the time the \gls{HLTDS} is being finalized, one should imagine demonstrating a series of steps on simulated imaging+prism data that will mimic what will be done with actual \gls{HLTDS} data:
\begin{enumerate}
\item Perform a calibration using simulated observations of wavelength and flux standards.
\item Having determined the calibration, simulate SN extractions from realistic imaging+prism data.
\item Assemble a simulated cosmology sample of SNe Ia with redshifts + light curves, investigate which SNe are getting misclassified photometrically (or assigned the wrong galaxy as the host) and improve the photometric classification.
\item Using both simulated spectra and simulated light curves, examine the population distributions of a set of SN parameters as a function of redshift and host-galaxy type, and see what evidence of population drift there is compared to lower redshift.
\item Select a volume-limited prism SN sample covering reasonable S/N, fit it with the \gls{SED} model, and do dimensionality reduction (e.g., principal component analysis) on the residuals to see if there is any evidence that the training of the \gls{SED} model has missed some behavior of SNe.
\item Finally, perform a simulated cosmology analysis using the above analysis products.
\end{enumerate}

The collection of studies presented in this work are an existence proof that the above series of steps is possible and will yield useful data; Section~\ref{sec:analysisoverview} outlines the analyses we present. Section~\ref{sec:prismsurveysim} presents a simple demonstration survey that (although this survey is not fully optimized) shows the performance of the prism that is possible for each tier in a \TierSlash \NumTiers-tier survey, examining S/N and numbers of SNe~Ia as a function of redshift. Section~\ref{sec:performance} outlines our evaluation of this survey according to the \numberpoints uses of spectroscopy above. These studies show that a prism can produce spectral time series with S/N sufficient for the goals I and III listed above (redshift measurements and \gls{SED} constraints), with conclusive results on goal II (typing) needing further study with a core-collapse SN time-series model. Section~\ref{sec:performance} also performs a simple survey optimization, maximizing the number of SNe~Ia useful for different purposes and forecasting the cosmological constraints possible with those SNe. Section~\ref{sec:performance} ends by describing the optimization of the prism parameters, and shows that the dispersion is high enough that extracting the data should be possible with only modest biases.
Finally, we conclude and provide a glossary in Section~\ref{sec:conclusion}. For the sake of readability, we put some of the technical discussion in appendices: Appendix~\ref{sec:prismsimulations} presents simulation details, Appendix~\ref{sec:analytic} shows a simple analytic optimization of the prism dispersion, and Appendix~\ref{sec:prismvsimaging} evaluates constraints on SN~Ia parameters in prism vs. imaging at fixed total survey time.

\begin{deluxetable}{cc|cc|cc}[htbp]
\rotate
\tablecaption{Overview of Analyses}
\label{table:analysisoverview}
\tablehead{
\colhead{Section(s)} & \colhead{Analysis} & \colhead{Simulation SED Model} & \colhead{Type of Simulation} & \colhead{Fitting SED Model} & \colhead{Type of Inference}}
\startdata
\ref{sec:prismsurveysim} & S/N, Numbers of SNe & SALT2-Extended & 1D Spectra & \nodata & \nodata \\
\ref{sec:redshift} & Redshift Measurements & SN Timeseries & 1D Spectra & SALT2-Extended & Minimization \\ 
\ref{sec:subclassification} & SN Ia Subclassification & SNEMO15 & 1D Spectra w + w/o Stacking & SNEMO15 & Fisher Matrix \\ 
\ref{sec:missingcomponent} & Missing SED Component & SNEMO15 & 1D Spectra & SNEMO15 & Minimization \\
\ref{sec:surveyoptimization} & Survey Optimization & SALT2-Extended & 1D Spectra & \nodata & Fisher Matrix \\
\ref{sec:prismoptimization} & Optimum Prism Parameters & SNEMO15 & Forward Model & SNEMO15 & Minimization \\
Appendix~\ref{sec:analytic} & Analytic Optimization & Gaussian Feature & 1D Spectrum & Gaussian Feature & Fisher Matrix \\
Appendix~\ref{sec:prismvsimaging} & Prism vs. Imaging Trade & SUGAR & 1D Spectra & SUGAR & Fisher Matrix \\
\enddata
\end{deluxetable}

\section{Overview of Our Analyses and Simulations} 

\subsection{Overview of Analyses} \label{sec:analysisoverview}

We perform a series of analyses to evaluate and optimize the performance of the prism; Table~\ref{table:analysisoverview} presents an overview of these analyses. Each analysis generally has two components: the model for simulating the spectra and the model for inference of results. As we show in Section~\ref{sec:prismsurveysim}, the prism can be used to build time series of SNe, so it is important to use full time-series models for both simulation and inference, and we use several different models depending on the goal. In general, having more models provides cross checks and increases the robustness of our results. We describe here some of the considerations for why we selected the models that we did.

\gls{SED} models for simulations:
\begin{itemize}
    \item The \gls{SALT2}-Extended model \citep{SALT2, Pierel2018} is a combination of \gls{SALT2} and the \citet{Hsiao2007} SN~Ia template. It spans the largest rest-frame wavelength range (1,000\Ang--18,000\Ang) of any of our \gls{SED} models. This makes it the best choice for spanning large redshift ranges, for example predicting S/N for the whole survey or fitting simulated data to find redshifts. However, \gls{SALT2} only has one intrinsic parameter of SN variability ($x_1$, which has no effect outside the rest-frame optical) and one color parameter ($c$), so it is not the best choice for looking at SN~Ia variability in detail.
    \item \gls{SNEMO} is an \gls{SED} model based on a principal-component decomposition of the \gls{SNfactory} spectral time series \citep{Saunders2018}. It only spans a rest-frame wavelength range of 3,300\Ang--8,600\Ang, so it can only fit some of the \WFIRST observer-frame wavelength range. It comes in versions with one (SNEMO2), six (SNEMO7), and fourteen principal components (SNEMO15) of variability (plus color). It is possible that these large numbers of linear principal components (especially SNEMO15) may be approximating nonlinear trends in the data with linear components\footnote{This may be related to SNEMO7 performing roughly comparable to \gls{SALT2} in standardization performance \citep{Rose2020}.}; \citet{Rubin2020} and \citet{Boone2021a} suggest that $\sim$~three intrinsic components may be closer to the right number.
    \item \gls{SUGAR} \citep{Leget2020} spans a rest-frame wavelength range of 3340\Ang--8580\Ang, comparable to \gls{SNEMO}. Its main advantage over \gls{SNEMO} is that it describes SNe~Ia with three intrinsic parameters (plus extinction).
    \item We also directly use SN time series for simulations. We draw on the spectrophotometric \gls{SNfactory} data for this \citep{Aldering2020}, which (at present) is available for a rest-frame wavelength range of 3,300\Ang--8,600\Ang.
\end{itemize}

We use two types of simulations:
\begin{itemize}
    \item 1D simulations simply produce spectra with proper wavelength sampling and S/N. Generally speaking, our analyses fit the time series of 1D spectra directly (without stacking). Note that we call these simulations ``1D,'' even though they produce a time series of 1D spectra and thus could be considered 2D (wavelength and time).
    \item Forward models produce 2D simulated images (which we generate assuming a four-point $2\times2$ dither pattern) which are optimally extracted using a forward model (e.g., \citealt{Bolton2010, shukla17, Ryan2018}). We only use this more computationally intensive approach when optimizing the prism parameters and looking at robustness to an inaccurate \gls{LSF}. Again, our analyses fit the full time series directly. Note that we call these simulations ``2D,'' even though they produce a time series of 2D images and thus could be considered 3D (along the trace in wavelength, perpendicular to the trace, and time).
\end{itemize}

Finally, we consider two types of inference:
\begin{itemize}
    \item We use least-squares minimization to find the best-fit parameters for some of our analyses. For fitting redshifts (a somewhat nonlinear process that generally produces local $\chi^2$ minima), we try many initial starting redshifts to ensure convergence to the global best-fit. When fitting 1D simulations (each epoch represented by a vector of fluxes and a vector of uncertainties), we use a spectral model to generate model values at those specific wavelengths and dates. The 2D simulations (simulated images) are also fit with least squares, with the forward-model code used as a generative model for each epoch of simulated data.
    \item We use Fisher-matrix calculations to predict uncertainties (the inverse of the Fisher matrix is the parameter covariance matrix) by linearizing a model in its parameters and approximating the observational uncertainties as Gaussian. We use Fisher-matrix calculations when using \gls{SED} models with limited rest-frame wavelength coverage (\gls{SNEMO} and \gls{SUGAR}). This limited wavelength coverage may require using only modest redshift perturbations around the true redshift to avoid shifting some wavelengths outside the wavelength range of the model, so a Fisher-matrix approach is best. Fisher-matrix calculations are also useful to quickly get uncertainty estimates (but as noted above, they are not fully trustworthy for assessing redshift measurements due to multiple $\chi^2$ minima).
\end{itemize}

We do not consider model-independent spectral-feature measurements (e.g., smoothing and splines in \citealt{Blondin2006}) for two reasons. 1) As we show in Section~\ref{sec:prismsurveysim}, individual visits with even $\sim 1$~hour exposure times only yield low S/N spectra which are only built up over multiple visits to reasonable time series. The SN spectral features evolve over this time, so a parameterized model is the most efficient and unbiased method of inference. 2) The optimal prism dispersion (discussed in Section~\ref{sec:prismoptimization}) samples SN spectral features well, but (unlike for observations with $R\sim1000$ ground-based spectrographs) does not over-resolve so much that the \gls{LSF} does not matter. Once again, a parameterized model (that can be convolved with the \gls{LSF}) will be necessary to perform unbiased inference.

\subsection{Prism Survey Simulation} \label{sec:prismsurveysim}

We present here a simple survey simulation that demonstrates the utility of a slitless prism on \WFIRST. We discuss the simulation details in Appendix~\ref{sec:prismsimulations}. Figure~\ref{fig:prismsens} shows our estimated point-source sensitivity of the prism.

\begin{figure}[htbp]
    \centering
    \includegraphics[width=0.95 \textwidth]{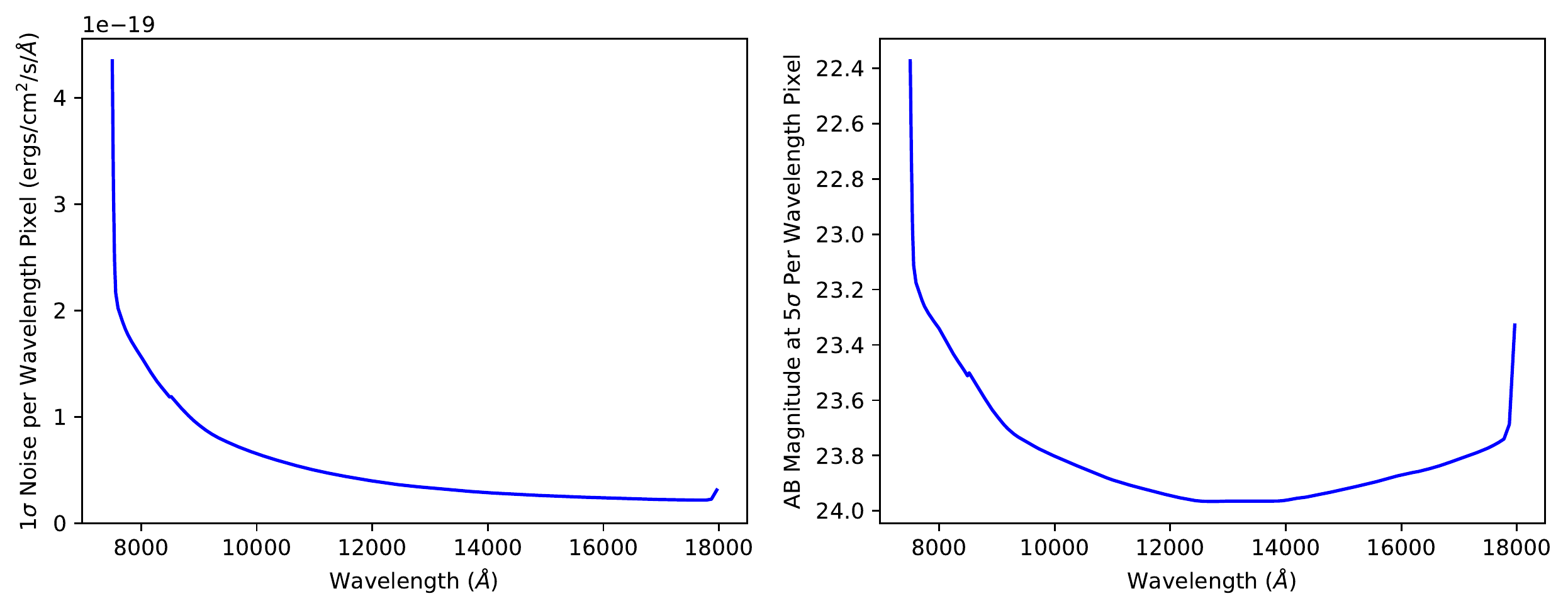}
    \caption{{\bf Left panel}: 1$\sigma$ noise level per pixel in wavelength for a one-hour prism exposure of a point source. {\bf Right panel}: The AB magnitude that can be measured at 5$\sigma$ (not 1$\sigma$), shown for each pixel in wavelength.}
    \label{fig:prismsens}
\end{figure}

Our simulated survey is similar to the \gls{SDT} survey \citep{spergel15} in that it envisions a two-year mission with 30 hour visits once every 5 days (2--5 days in the rest frame, depending on redshift), with a total integrated time of 6 months (including overheads). Optimizing the split between imaging and prism spectroscopy will be a focus of future work. For now, we simulate a 25\% prism survey (cf. the 10\%, 25\%, 50\%, and 75\% in \citealt{Rose2021}). One can roughly scale our numbers of SNe by the fraction of time used for the prism (i.e., half the time for the prism means that our numbers should be doubled).

The \NumTiers prism tiers consist of a \widedegsq square degree (\widepointings) wide field with an exposure time of \wideexp~s and a \deepdegsq square degree deep field (\deeppointings) with an exposure time of \deepexp~s. Including slew times of \slewtime, every five days \widehours hours are spent on the wide tier and \deephours hours are spent on the deep tier. We assume a simple four-point dither for each visit. The longer deep-tier exposures are Poisson-dominated and thus could certainly be broken into more dithers without taking a read-noise penalty, but this is only a minor effect for our simulations.

\begin{figure}[htbp]
    \centering
    \includegraphics[width=0.6 \textwidth]{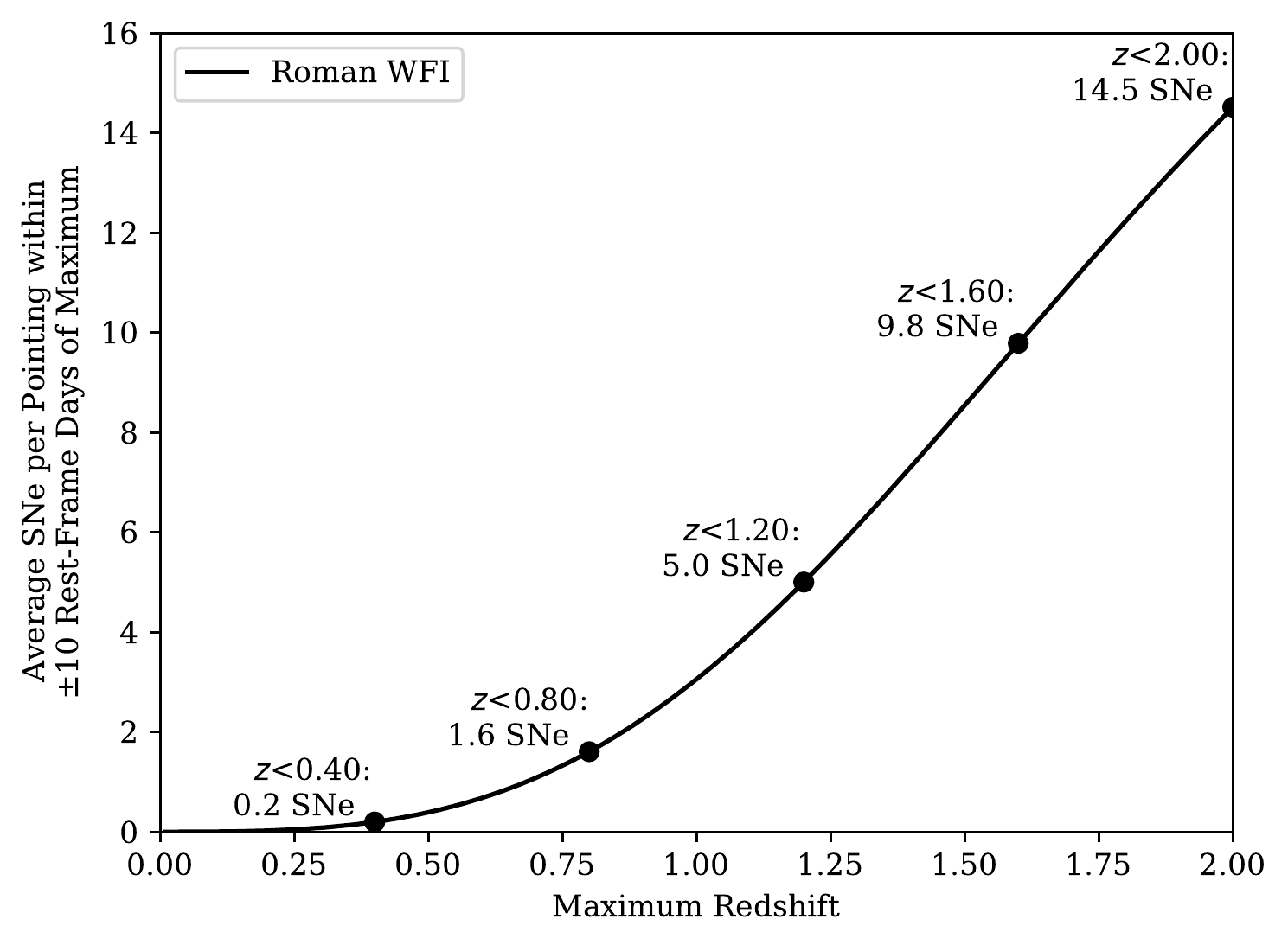}
    \caption{\gls{WFI} multiplex: average number of SNe~Ia per pointing within $\pm$10 rest-frame days of maximum vs. redshift, updated from \citet{Rose2021}. Assuming that the \WFIRST SN field(s) are in the continuous-viewing zone so that survey temporal edge effects from stopping and starting are small, the multiplex for other phase ranges scales from this plot (e.g., the multiplex within $\pm5$ rest-frame days is half the plotted values). Multiplex increases rapidly with maximum redshift and is a significant factor in survey design for surveys with maximum redshift $\gtrsim 0.7$. Additional multiplex is possible for cadenced, untargeted survey strategies as SN-free (``reference'') observations do not have to be taken.
    }
    \label{fig:multiple}
\end{figure}

\begin{figure}[htbp]
    \centering
    \includegraphics[width=0.5\textwidth]{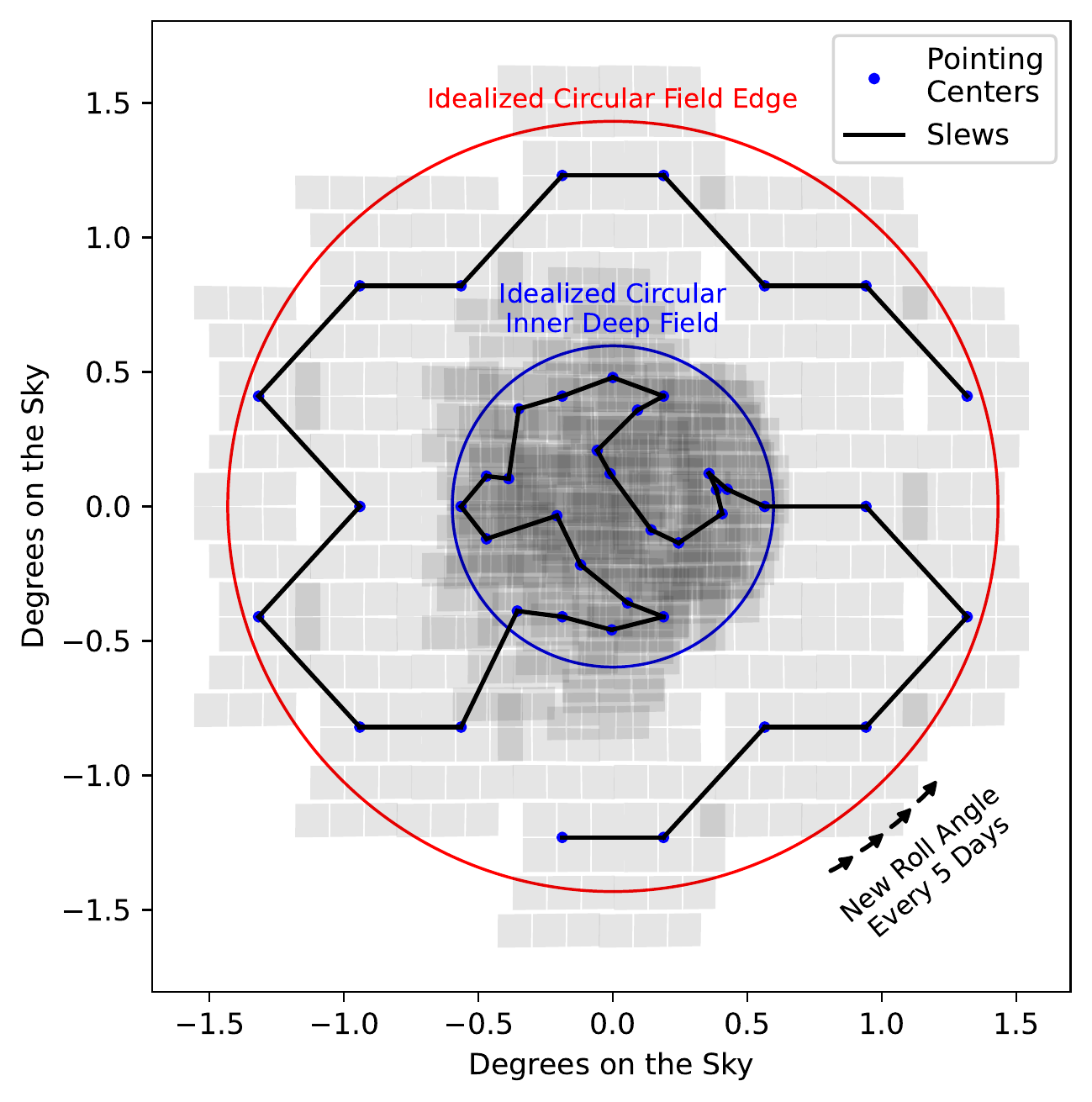}
    \caption{The idealized circular survey geometry (with no edge effects) assumed in this work overlaid with a realistic survey geometry made from \gls{WFI} pointings (shaded gray). The blue dots mark the centers of \gls{WFI} pointings, and the black lines connect these centers with the shortest total slew path (found with Concorde, \citealt{Concorde}). For illustrative purposes, we do not show chip-gap-spanning dithers. Each cadence step (assumed in this work to be five days), \WFIRST rotates to keep its solar panels pointing at the Sun. These 72 discrete roll angles have the fortunate effect of helping to break spatial/spectral degeneracy in the slitless prism spectroscopy.}
    \label{fig:pointings}
\end{figure}

The prism spectroscopy covers the full \SqDegSlash areas in each visit. In this way each prism exposure will frequently contain multiple live SNe (shown in Figure~\ref{fig:multiple}), with varying signal to noise per spectrum. As Figure~\ref{fig:multiple} shows that the multiplex factor is already significant by $z \sim 0.8$, we do not consider targeted prism observations of lower-redshift SNe in this work. Figure~\ref{fig:pointings} shows that the wide and deep tiers cannot be exactly circular, so they will have to be embedded together (deep inside wide) into a larger (roughly circular) area.\footnote{We choose the pointing centers for the wide tier with a simple algorithm that fills a circular area, column by column. The deep tier is more complicated. For simplicity, we choose to implement each of the four 3600~s deep pointings as twenty three 600~s wide pointings (with one pointing going to the wide to enable a symmetric pointing pattern). The positions of four of these pointings are fixed to continue the wide-tier pattern. The positions of the other nineteen are chosen with a downhill simplex code \citep{NelderMead} that optimizes the actual depth on the sky compared to a uniformly filled circular field. After solving for all field centers, we solve for the path of shortest slew time with Concorde \citep{Concorde}. Concorde solves the traveling salesman problem, and can do so with non-Euclidean distances between points. For the path shown here, we use a table of slew times as a function of angular size to construct a matrix of slews between every set of points. As the traveling salesman problem solves for a cycle, we have a virtual point that is zero distance from all others, then remove this point to obtain a solution where the starting and ending points are not the same.} This will necessarily blend SNe from the tiers together, as (a fraction of the) SNe rotate between them as \WFIRST rotates throughout the year to keep its solar panels pointed at the Sun. We simulate the tiers as though they are completely distinct, as this has sufficient accuracy for our purposes here.

\begin{figure}[htbp]
    \centering
    \includegraphics[width=0.6 \textwidth]{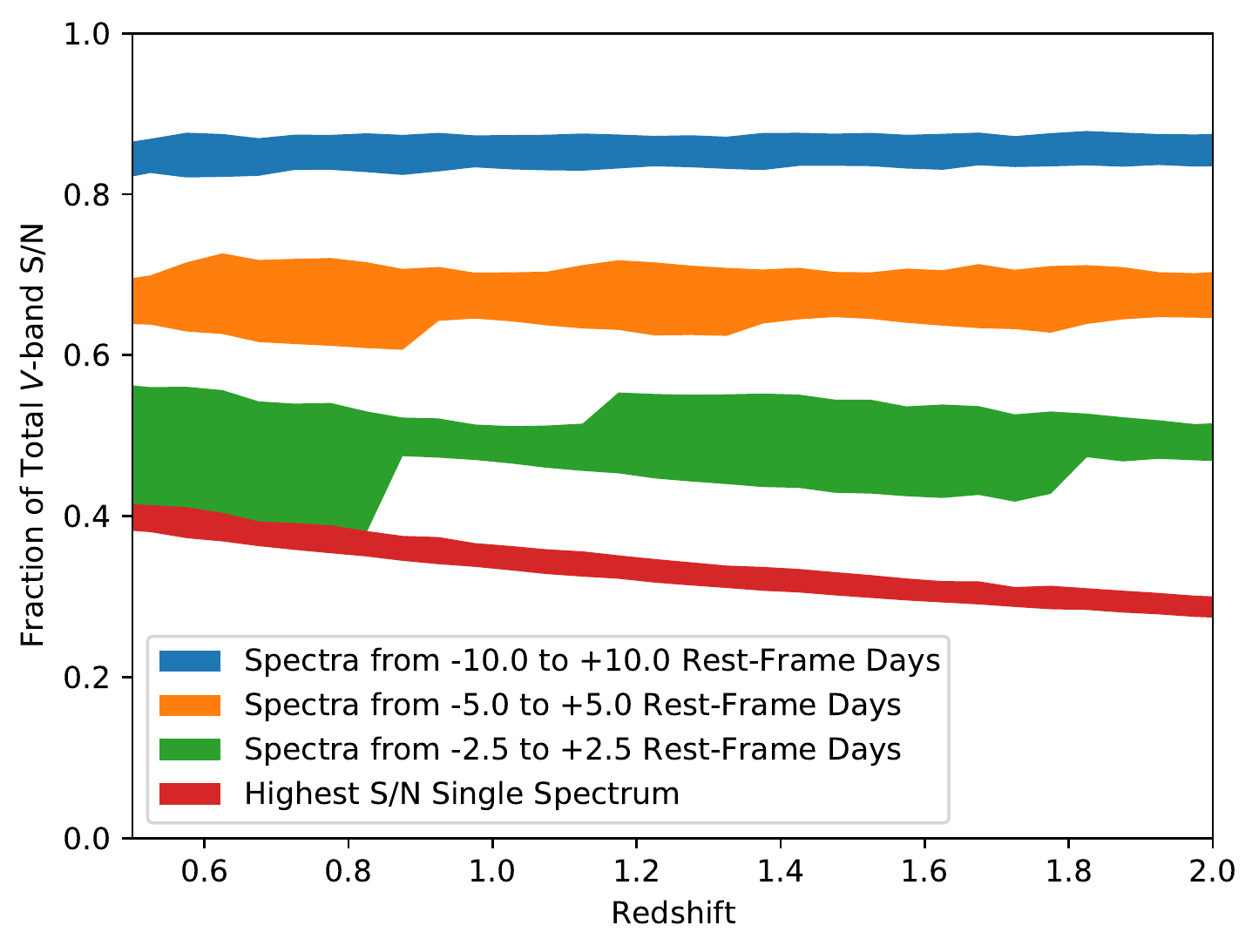}
    \caption{Fraction of the total time series rest-frame $V$-band S/N in spectra around maximum for SNe~Ia. The red confidence interval (encompassing 68.3\% of the SNe) shows the fraction of the S/N in the single highest S/N spectrum. This declines as $\sqrt{1 + z}$, as the fixed rest-frame windows encompass more spectra at higher redshifts due to time dilation, but is generally around 1/3rd (e.g., a S/N 75 SN in Figure~\ref{fig:SNhistNew} has a $V$ S/N of around 25 for the spectrum closest to maximum light). The green interval shows the fraction of the S/N integrating from $-$2.5 to +2.5 rest-frame days. Quantization is visible in certain redshift ranges, e.g., at redshift 0.5, the rest-frame cadence is $3.3\overline{3}$ days, so sometimes a 5 rest-frame-day window contains two spectra, and sometimes just one. Orange and blue show $-$5 to +5 and $-10$ to +10, respectively. Analyses that efficiently use time-series information will be necessary to reach the full potential of the prism data.}
    \label{fig:SNRmax}
\end{figure}

\begin{figure}[htbp]
    \centering
    \includegraphics[width=0.49 \textwidth]{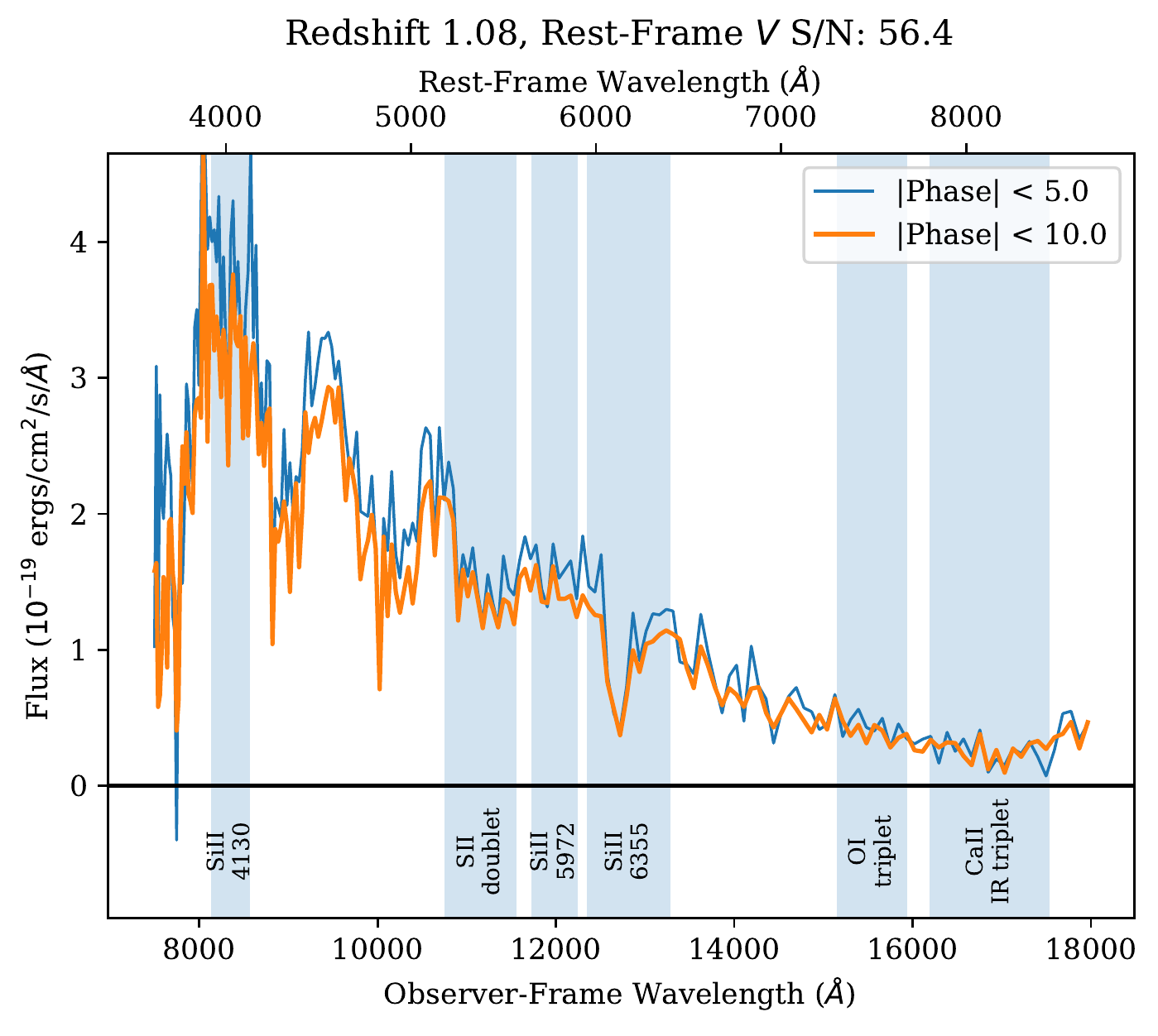}
    \includegraphics[width=0.49 \textwidth]{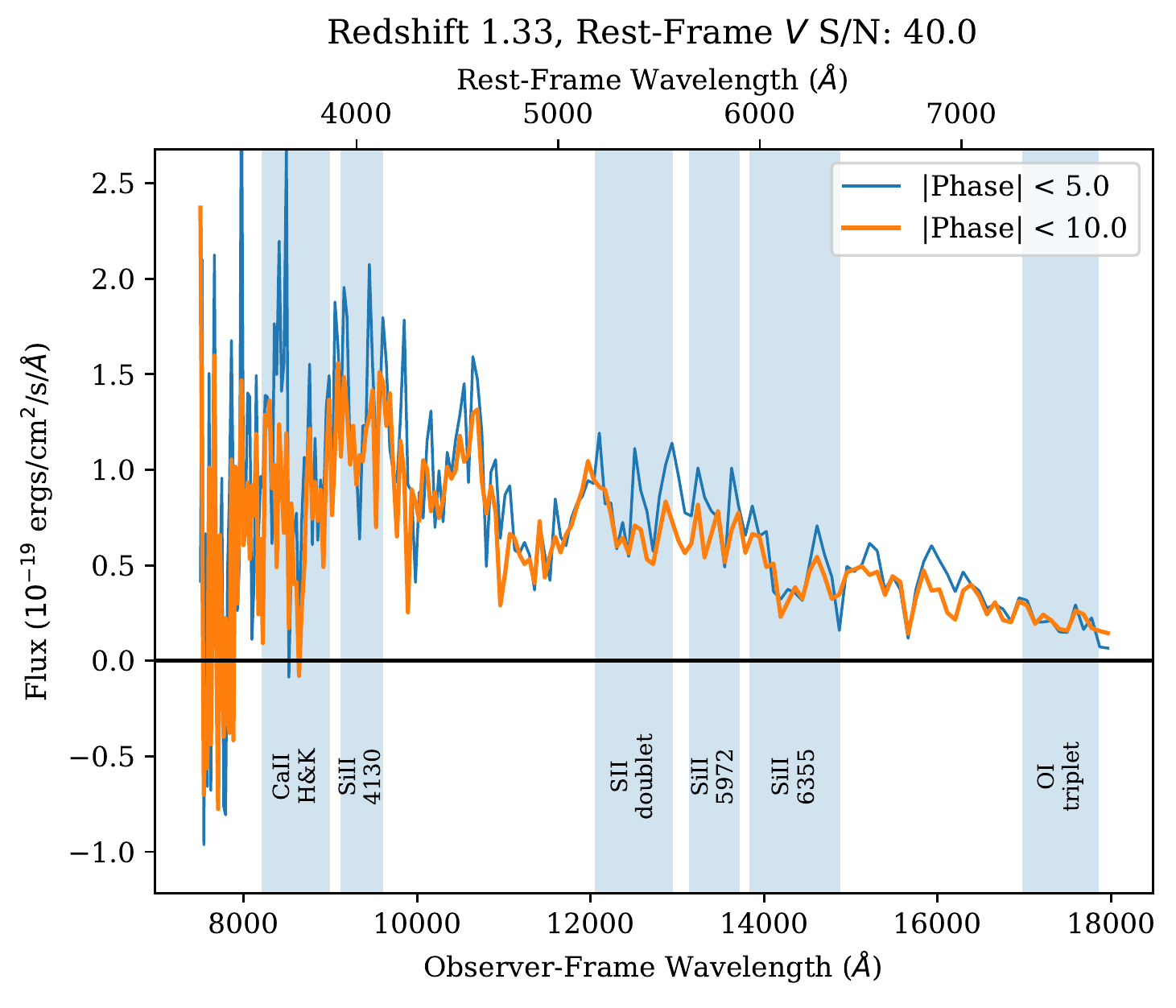}
    \includegraphics[width=0.49 \textwidth]{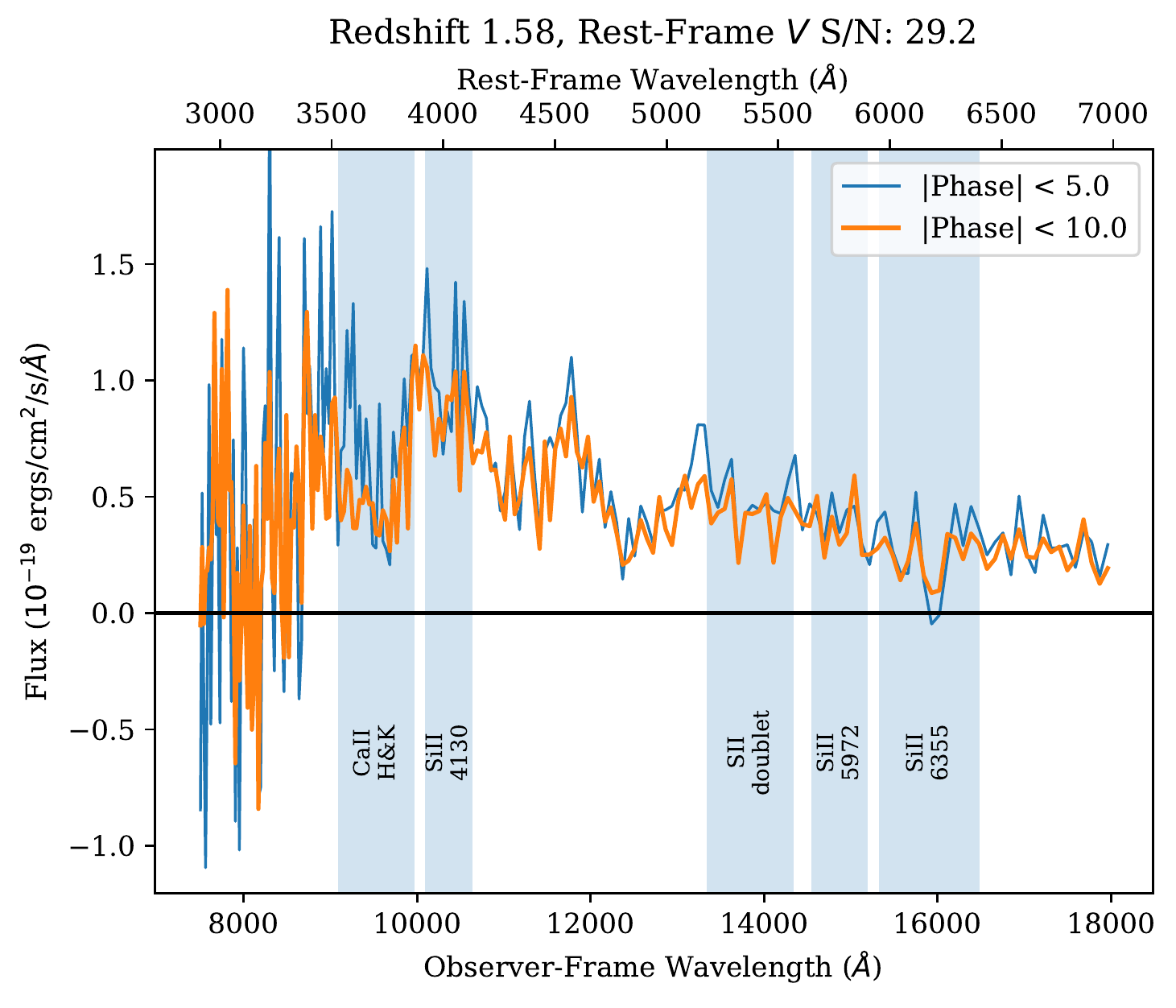}
    \includegraphics[width=0.49 \textwidth]{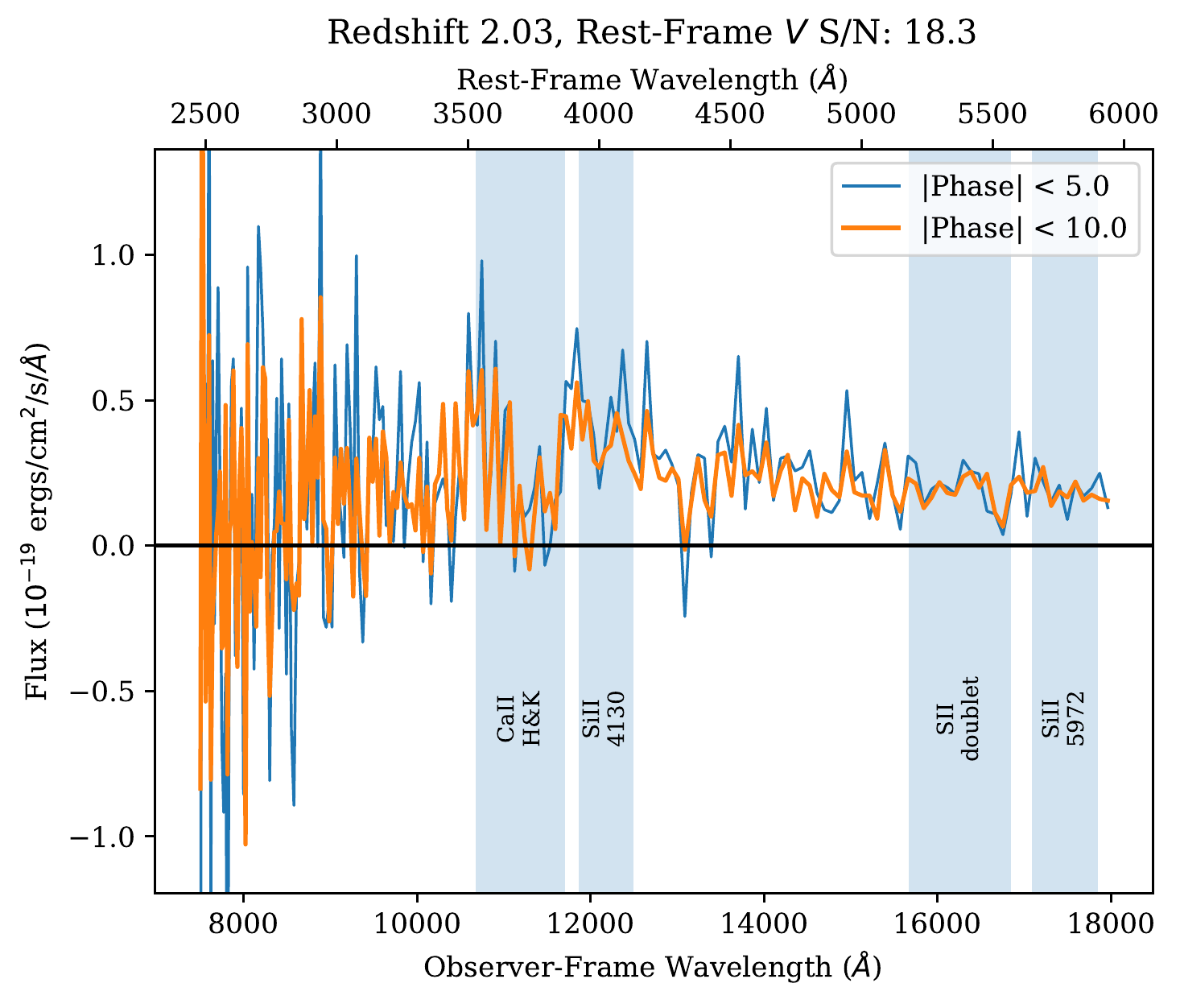}
    \caption{Simulated SNe Ia with one hour per visit every five days. Each SN shown has the median S/N out of many simulated at its redshift. We stack the data within five and ten rest-frame days of maximum for visualization purposes (Section~\ref{sec:subclassification} shows that stacking is not an optimal analysis). The wavelengths shown are the native wavelengths of the prism with no smoothing. Redshift $\sim 1.1$ SNe~Ia ({\bf top left}) are very clearly SNe Ia, and subclassification is possible. Redshift $\sim 1.3$ SNe~Ia ({\bf top right}) can generally be recognized as SNe Ia (at a minimum, the Si~6355\,\Ang and 4130\,\Ang features are visible), and probabilistically subclassified. At redshift $\sim 1.6$ ({\bf lower left}), classification may be possible. Finally, at redshift $\sim 2$ ({\bf lower right}), the $\sim 2800$\Ang break and Ca H\&K are visible (with lower S/N for other features), but the strong Si 6355\,\Ang feature is not covered (and would not have enough S/N for a strong detection, even if the prism went redder). This SN is thus recognizable as a $z=2$ SN~I, but probably cannot be absolutely classified as a SN~Ia (cf., \citealt{rubin13, Jones2013}).
    \label{fig:simspectra}}
\end{figure}

We need a way to summarize the S/N of a spectral time series into one number that can reasonably represent the useful S/N if, e.g., the cadence or dispersion changes. We choose to compute the combined S/N of all spectra, integrated over a tophat from 5000\AA\xspace to 6000\AA\xspace in the rest frame (hereafter referred to as the ``$V$'' band):
\begin{equation}
    V_{\mathrm{SNR}} \stackrel{\text{def}}{=} \sqrt{ \sum_{\mathrm{Spectrum}\ i} \ \  \sum_{\mathrm{Wavelength}\ j \in V} \left[ \frac{\mathrm{True\ Noiseless\ Flux}_{ij}}{\mathrm{Uncertainty}_{ij}} \right]^2 }
\end{equation}
The rest-frame $V$ is a reasonable choice: it is accessible for most of the relevant redshift range and it is adjacent to the strong Si 6355\,\Ang feature (which is blueshifted around maximum light to $\sim 6150$\,\Ang). Note that the S/N values listed here are thus, for convenience of comparison, based on the total-time-series SN spectra that would be obtained over a period of time, though all of the analyses we have seriously considered would fit the time series of individual spectra (or an individual spectrum near max), not the co-added stack. Not all of these spectra would be around maximum light and therefore their contribution to the total-time-series S/N would be less significant. The contribution of each spectra within a given time frame from peak to the overall S/N, is presented in Figure~\ref{fig:SNRmax}. Figure~\ref{fig:simspectra} shows simulated SN observations in increasing redshift order. 

\begin{figure}[htbp]
    \centering
    \includegraphics[width=0.45 \textwidth]{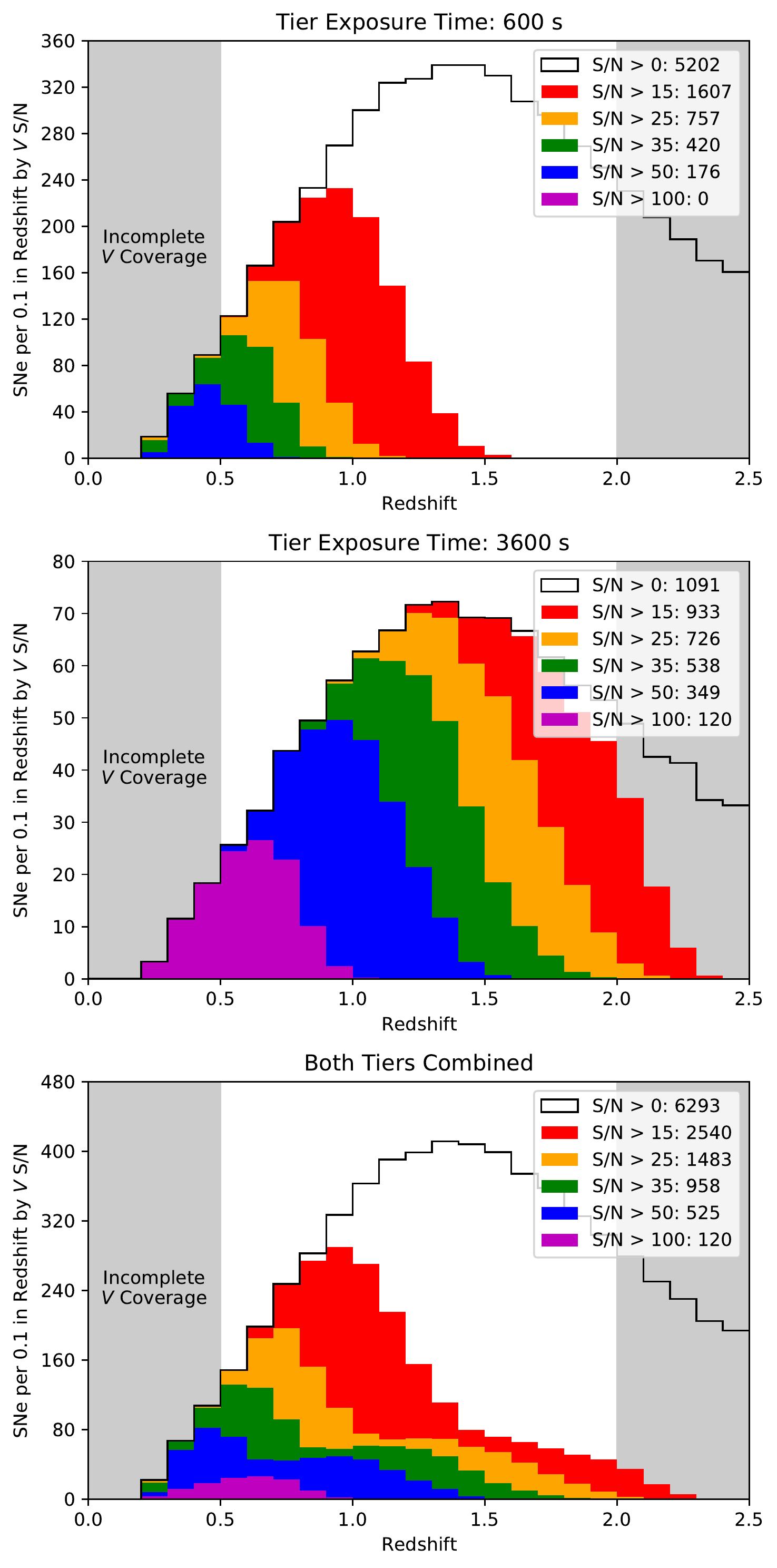}
    \includegraphics[width=0.45 \textwidth]{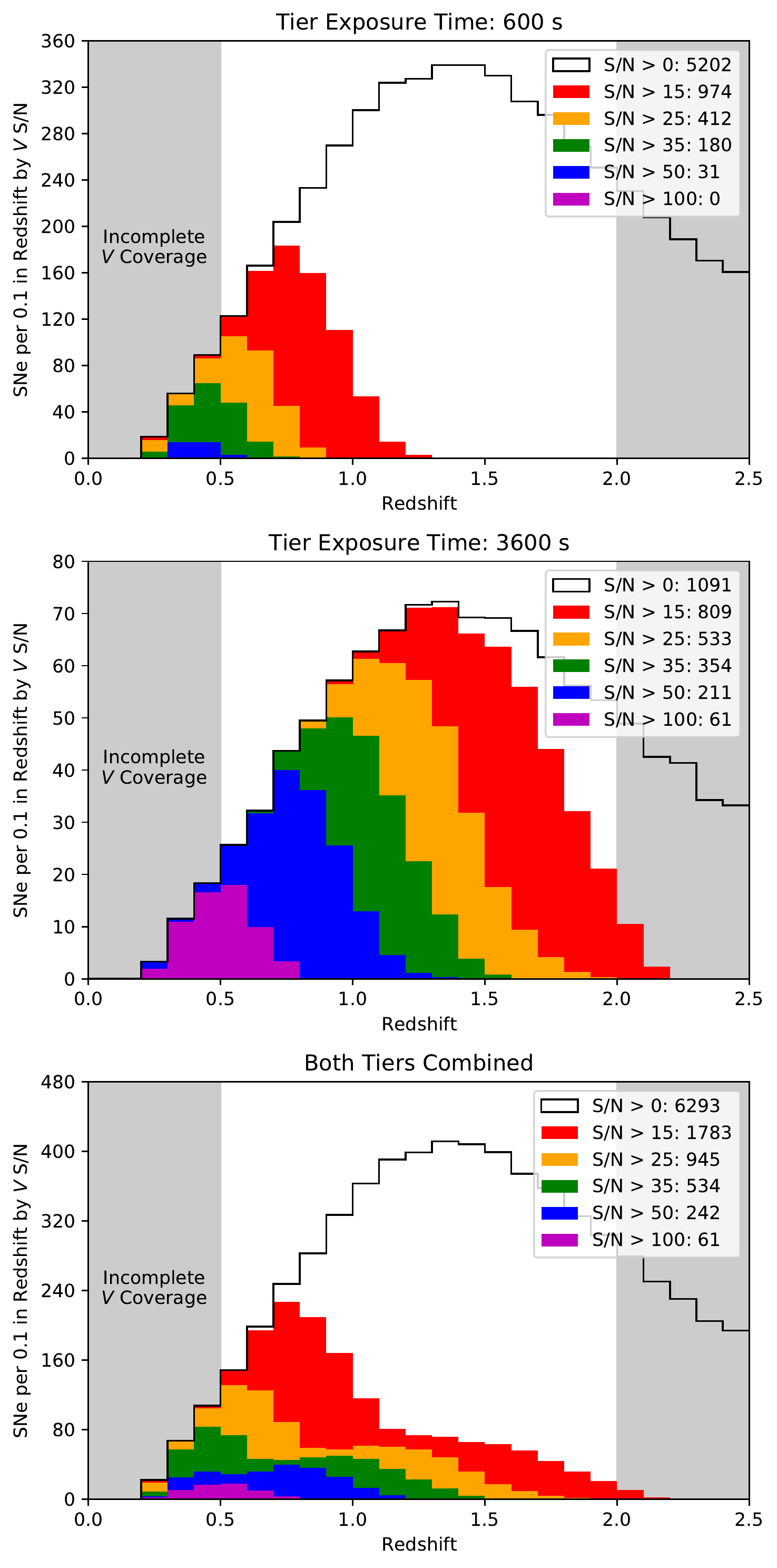}
\caption{SN~Ia yields from our demonstration survey (we average over 25 survey realizations to reduce Poisson noise). {\bf Left column}: numbers of SNe assuming 3D host-galaxy subtraction (or no need of host subtraction). {\bf Right column}: numbers of SNe taking a 2D $\sqrt{2}$ host subtraction noise hit, in which only SN-free references at a given rotation angle are used. The numbers of SNe in the right column are high enough to be useful, but the gains from a 3D host-galaxy subtraction are clear. For the time series of each SN, we compute the total rest-frame $V$-band signal to noise of the full time series as described in Section~\ref{sec:prismsurveysim} and color code accordingly. We shade the redshifts where the rest-frame $V$ band wavelength range is incompletely covered by the prism in gray.}
    \label{fig:SNhistNew}
    \end{figure}

Figure~\ref{fig:SNhistNew} shows the number of SNe within a certain redshift bin color coded by S/N. Within this figure the \TierOne prism tier data is shown in the top panel, the \TierTwo tier below that, and the sum of the tiers on the bottom. The left column shows the numbers of SNe assuming a SN-free (``reference'' or ``template'') observation much deeper than the live-SN observations (thus not contributing any correlated noise to the SN time-series when subtracting host-galaxy light). This will require combining reference observations taken over two complete rolls (as the \gls{HLTDS} survey is planned to last for two years and \WFIRST must keep its solar panels pointed near the Sun). Each roll angle will sample the spatial and spectral information of nearby objects differently. We thus refer to this as a ``3D'' host-galaxy subtraction, as a 3D model (coordinates on the sky and wavelength) will have to be constructed for the part of the sky that can blend into the spectral time series of each SN. The right column shows the case where observations only at the same roll angle can be used to subtract the host-galaxy light, so the observations decrease in S/N by a factor $1/\sqrt{2}$.

In total, Figure~\ref{fig:SNhistNew} shows that we expect to obtain the redshifts of $\sim \NSNefifteen$ Type Ia SNe ($\sim \NSNetwentyfive$ with secure types and redshifts). Of these, \NSNethirtyfiveHost--\NSNethirtyfive will have S/N $> 35$, be sub-typed, and be suitable for population and evolution studies. Finally, \NSNefiftyHost--\NSNefifty SN Ia will have S/N $> 50$ and will be used for the retraining of the \gls{SED} model at high redshift. Again, these numbers are for a survey with 25\% time devoted to the prism (0.125~years of prism time), and roughly scale linearly with that amount.

\clearpage

\section{Evaluating the Scientific Performance of a Prism} \label{sec:performance}
The simulated survey strategy we present is based on the simple optimizations in Section~\ref{sec:surveyoptimization}. Although future work will improve on this study, this survey does however provide a baseline from which to evaluate each of our scientific goals. We discuss the results below, focusing on the results for the deep tier (\deepexp~s of exposure time per visit), as there is the best overlap between the prism wavelength coverage and the current set of (rest-frame-optical-focused) models for $z \gtrsim 1$.

\subsection{Redshift Measurements} \label{sec:redshift}

\begin{figure}[htbp]
    \centering
   \includegraphics[width=0.95\textwidth]{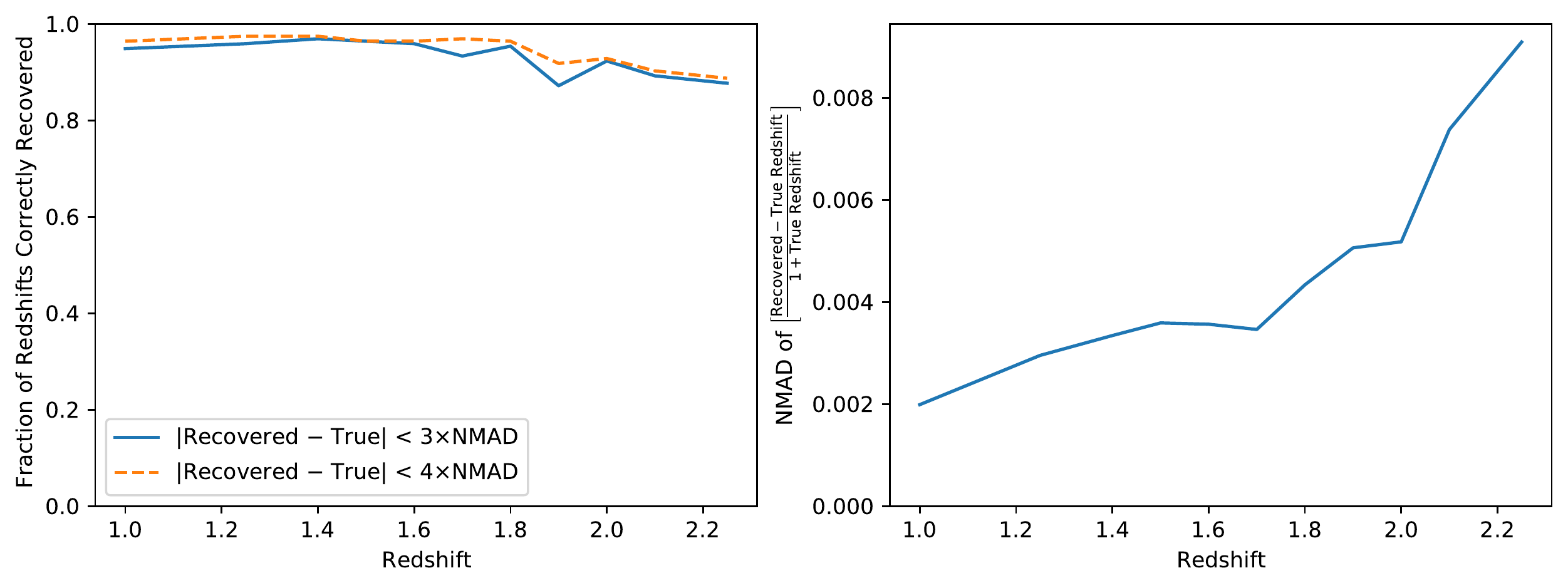}
   
   \includegraphics[width=0.95\textwidth]{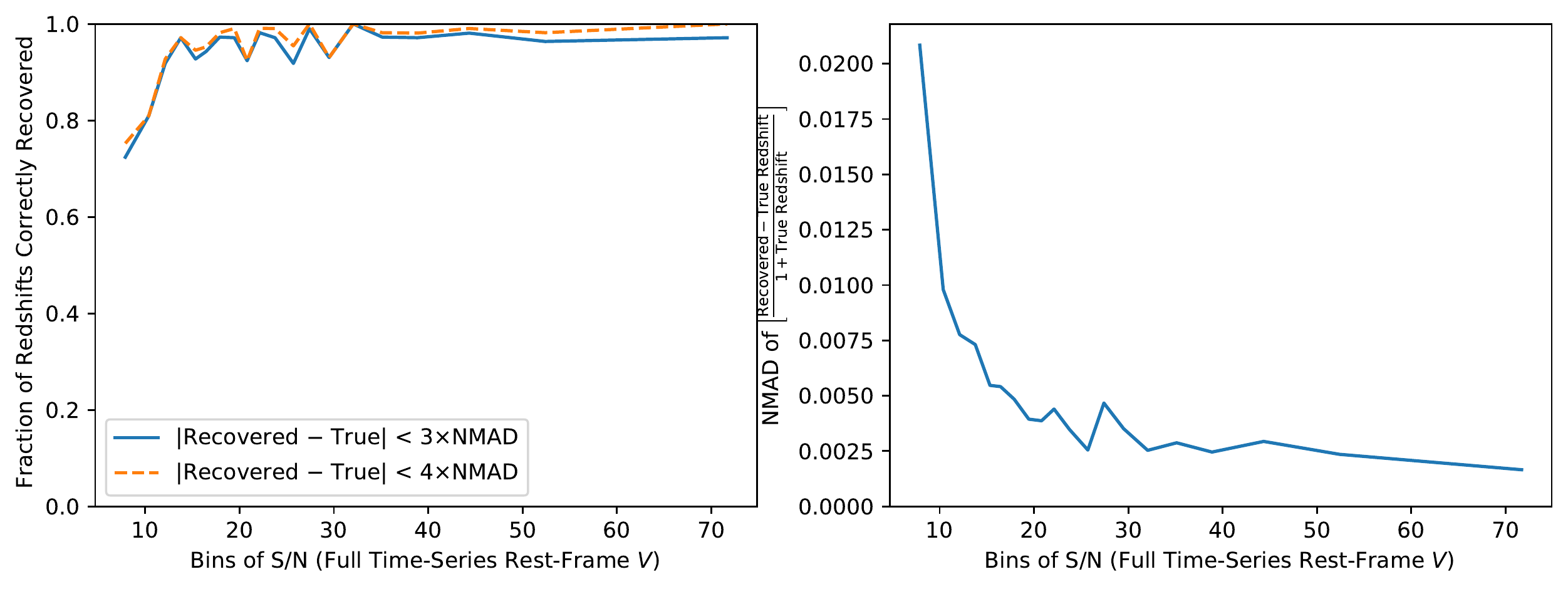}
   
    \caption{The {\bf top left panel} shows the fraction of our simulated SN~Ia prism spectroscopic sample that is recovered near the core of the distribution as a function of simulated SNe redshift. We take the spectral time-series data from the 39 SNe in the Nearby Supernova Factory data set \citep{Aldering2020} having the best temporal sampling, and degrade the wavelength sampling and signal-to-noise to match 1-hour \WFIRST prism exposures. Each SN is simulated five times, with five different realizations of noise. To recover the redshifts, we fit the \gls{SALT2}-extended spectral model to the simulated prism data, using a wide range of initialization redshifts to ensure convergence to the global best fit. The plateau in efficiency at low redshift is due to a peculiar 1991T-like SN \citep{Filippenko1992, Phillips1992} in our input data that is not well modeled by \gls{SALT2}. Such SNe are modeled well by, e.g., the \citet{Boone2021a} parameterization, but we leave this for future work. Another future improvement on these results would come from simulating a dataset with bluer rest-frame wavelength coverage, as SNe Ia show a rapid decline in flux in the rest-frame NUV that can yield a secure redshift in this redshift range \citep{rubin13}. The {\bf top right panel} shows the dispersion (Normalized Median Absolute Deviation) in recovered redshifts. The {\bf bottom panels} repeat these summaries, but show the results in bins of rest-frame $V$ S/N. In general, these tests show reasonable performance to $z \lesssim 2$ and S/N $\gtrsim 15$.}
    \label{fig:completeness-snf}
\end{figure}

To assess the viability of measuring SN~Ia redshifts with the prism, we simulate time series and fit with \gls{SALT2}-extended. Using a model with wide rest-frame wavelength coverage enables us to initialize the optimization at a wide range of redshifts to ensure the redshift found really is the global optimum (in general, redshift finding can produce local $\chi^2$ minima, especially at low S/N, when the spectral features are hard to uniquely identify). We simulate directly from the \gls{SNfactory} time series \citep{Aldering2020}, using the 39 SNe with the base phase coverage. We interpolate each SN time series to a five-day observer-frame cadence, resample to prism resolution, and add appropriate noise for one-hour visits at each simulated redshift. Each SN is simulated five times, with five different realizations of noise. These time series do not cover the rest-frame UV, where the rapid falloff in flux can help secure the redshift \citep{rubin13}. Moreover, the \gls{SALT2}-extended model used here cannot fit a peculiar 1991T-like SN \citep{Filippenko1992, Phillips1992}, limiting the recovery rate, even at the lower redshift end of our simulations. A parameterization like \citet{Boone2021a} does fit 1991T-like SNe, and may do better if this parameterization can be extended into the UV. For both these reasons, our results should be considered an underestimate of the possible prism performance. Figure~\ref{fig:completeness-snf} shows the results of our simulated redshift measurements. One-hour exposures every five days are sufficient for measuring redshifts to $z \sim 2$ ($V$-band S/N $\sim 15$), a redshift that is difficult to reach from the ground.

\subsection{Typing of SNe} \label{sec:typing}

Most spectroscopic typing tools are based on a comparison of an unknown-transient spectrum to a set of library spectra, (e.g., \gls{SNID}, \citealt{blondin07}). We have experimented with using \gls{SNID} on our simulated time series to see if a majority of the spectra are classified as one type of SN. We see plausible results:  most simulated SNe~Ia are classified as SNe~Ia for most of their epochs with declining classification efficiency as the S/N varies from $\sim 35$ to $\sim 20$. However, the majority of training spectra (and spectral time series) are SNe~Ia, so a careful study of biases would need to be undertaken to come to a secure conclusion. A more promising path forward may be to use the \gls{ParSNIP} model \citep{Boone2021c}, which has a parameterization that spans both core-collapse and SNe~Ia. \gls{ParSNIP} has only been trained so far with imaging data, so we leave this for the future and suggest that rest-frame $V$-band S/N $\sim 25$ is necessary for typing (midway between the $\sim 15$ required for redshifts, and the $\sim 35$ required for sub-typing).

\subsection{Sub-typing of SNe Ia} \label{sec:subclassification}

In general, SNe~Ia are continuously distributed in most parameters \citep[e.g.,][]{Branch2006, Boone2021a}. We thus use ``sub-typing'' in this work to refer to placing SNe into regions of parameter space that are much smaller than the population distribution to obtain smaller distance uncertainties than are possible with a two-parameter \citet{tripp} standardization \citep{Wang2009, twins15, Boone2021b}. We take S/N $\sim 1$ (measurement uncertainty comparable to the population dispersion) as the threshold for useful sub-typing, which may seem like only a weak constraint. However, the \citet{twins15} analysis seems to be only S/N $\sim 1$ \citep{Rubin2020}; furthermore Bayesian Hierarchical Models can provide useful constraints on standardization coefficients and population parameters even when individual SNe are measured to this precision \citep{minka99, Hayden2019}.

\begin{figure}[htbp]
    \centering
    \includegraphics[width=\textwidth]{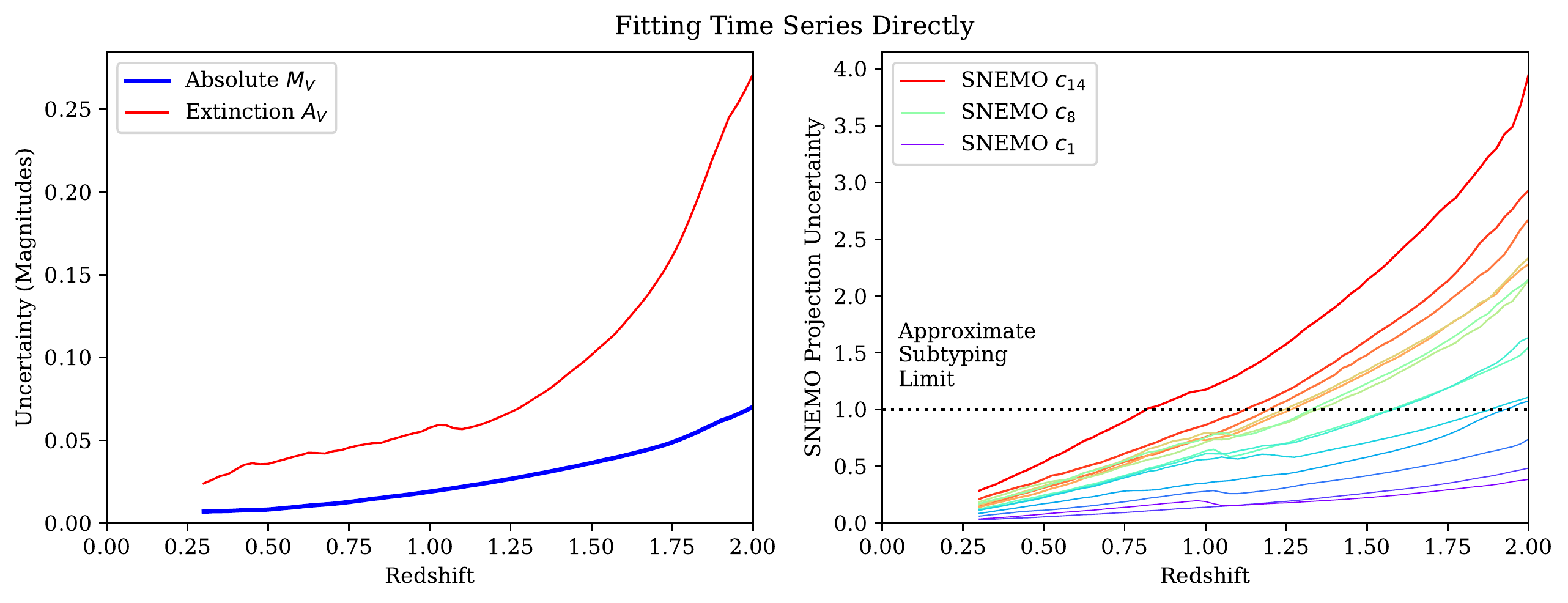}
    \caption{Prism time series (with one hour per visit) are simulated and fit with SNEMO15. The {\bf left panel} shows the absolute rest-frame $V$-band magnitude and extinction uncertainties; the {\bf right panel} shows the uncertainties on the coefficient for each eigenvector. (There are no published standardization coefficients for SNEMO15, so we cannot show distance-modulus uncertainties.) Generally, the SNEMO15 eigenvector projections are usefully measured to \redshiftsubtype, indicating that the prism time series can be used to sub-type SNe Ia.}
    \label{fig:prism_subtype}
\end{figure}

\begin{figure}[htbp]
    \centering
    \includegraphics[width=\textwidth]{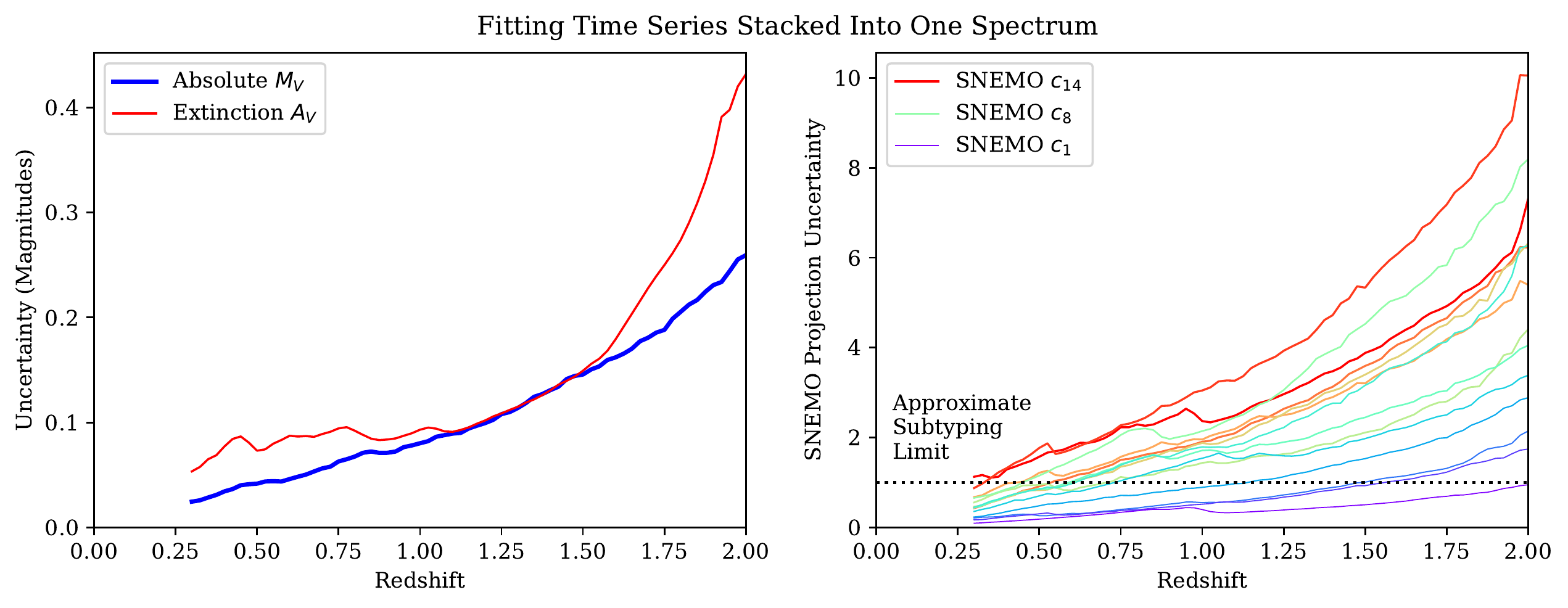}
    \caption{As in Figure~\ref{fig:prism_subtype}, but the time series is stacked using inverse-variance weighting into one spectrum before fitting with SNEMO (the SNEMO model is also stacked with the same weighting). As the time-series information is lost, this analysis includes a prior of $\pm 1$ day for the date of maximum. The uncertainties are much larger than the time-series analysis in Figure~\ref{fig:prism_subtype}, indicating that ignoring the time evolution with stacking throws away useful information (and that this is true even though each individual epoch is quite noisy).
    \label{fig:prism_subtype_stack}}
\end{figure}

 To determine the maximum redshift for which we can sub-type, we fit the 15-eigenvector \gls{SNEMO} model to simulated prism time series with 1 hour per visit.  Figure~\ref{fig:prism_subtype} shows the \gls{SNEMO}15 uncertainties as a function of redshift. The left panel shows the uncertainty in the rest-frame $V$ magnitude and the extinction $A_V$. The right panel shows the uncertainty on the scaling of each eigenvector. It shows a general trend where higher-numbered coefficients (e.g., $c_{14}$) have generally larger uncertainties than lower-number coefficients (e.g., $c_1$) at any given redshift. The higher numbered eigenvectors make much smaller contributions to the overall variation of SN fluxes and thus their projections require higher S/N to constrain. Figure~\ref{fig:prism_subtype_stack} shows the same analysis, but stacking the time series into one spectrum. The much larger uncertainties indicate that a time-series has much more information than a stacked spectrum at the same total S/N.

Most of the uncertainties in eigenvector projections cross our S/N $\sim 1$ threshold at \redshiftsubtype with one-hour exposures, when the time-series has a total S/N of $\sim 35$. We thus take this as a reasonable criterion for subtyping. A time-series model with standardization coefficients will be necessary to predict distance-modulus uncertainties as a function of redshift, which we unfortunately leave to future work.

\subsection{Population evolution and SED Training} \label{sec:missingcomponent}

The prism can provide time series of a random sample of SNe~Ia. These time series can investigate new, unknown SN behaviors, described as eigenvectors in a \gls{SNEMO}-like framework.\footnote{A similar concept is the intrinsic-scatter matrix \citep{Kessler2013}, which can be thought of as the sum of the outer products of the eigenvectors, with each eigenvector scaled by the width of the population distribution. We pursue an eigenvector-based description here because it is not clear that the population distribution will be the same at all redshifts and eigenvectors provide a natural basis for describing any changes in the mean SN.} We thus examine our ability to recover a eigenvector that is only present in high-redshift prism data. This eigenvector could represent the effect of a physical parameter that begins to be located outside the range it is observed to have at low redshift (i.e., where \gls{SNEMO} is trained). It may also represent an effect that shows up (partially or completely) outside the rest-frame wavelength range of \gls{SNEMO}.

We simulate 1 hour per pointing exposures of just 100 SNe at redshift 1 or 1.2. The time series for each SN is drawn from the SNEMO15 coefficient distribution in \gls{SNfactory} data. We get a baseline set of eigenvector projections for each SN by fitting its time series with SNEMO15, assuming knowledge of all 15 eigenvectors (the mean plus 14 components of variation).

Then, we pretend that we do not have knowledge of one of the eigenvectors, and recover this unknown eigenvector using an \gls{EMPCA} \citep{EMPCA}. This algorithm alternates between estimating projections (the ``e-step'') with estimating the missing eigenvector(s) (the ``m-step''). We initialize the missing eigenvector with 2D random Gaussian noise (the two dimensions are wavelength and phase, as with the other eigenvectors).\footnote{To impose some regularization on the solution, we use a 2D 3rd-order spline as our basis, with eight nodes in phase and 20 in wavelength. Examining other (combinations of) missing eigenvectors and regularization is left for future work.} Then we estimate all the eigenvector projections, and with the projections fixed, solve for the missing eigenvector. Using this updated eigenvector, we reestimate all of the projections, and iterate until convergence (defined as no eigenvector projection changing by more than 0.01).

\begin{figure}[htbp]
    \centering
    \includegraphics[width=0.65 \textwidth]{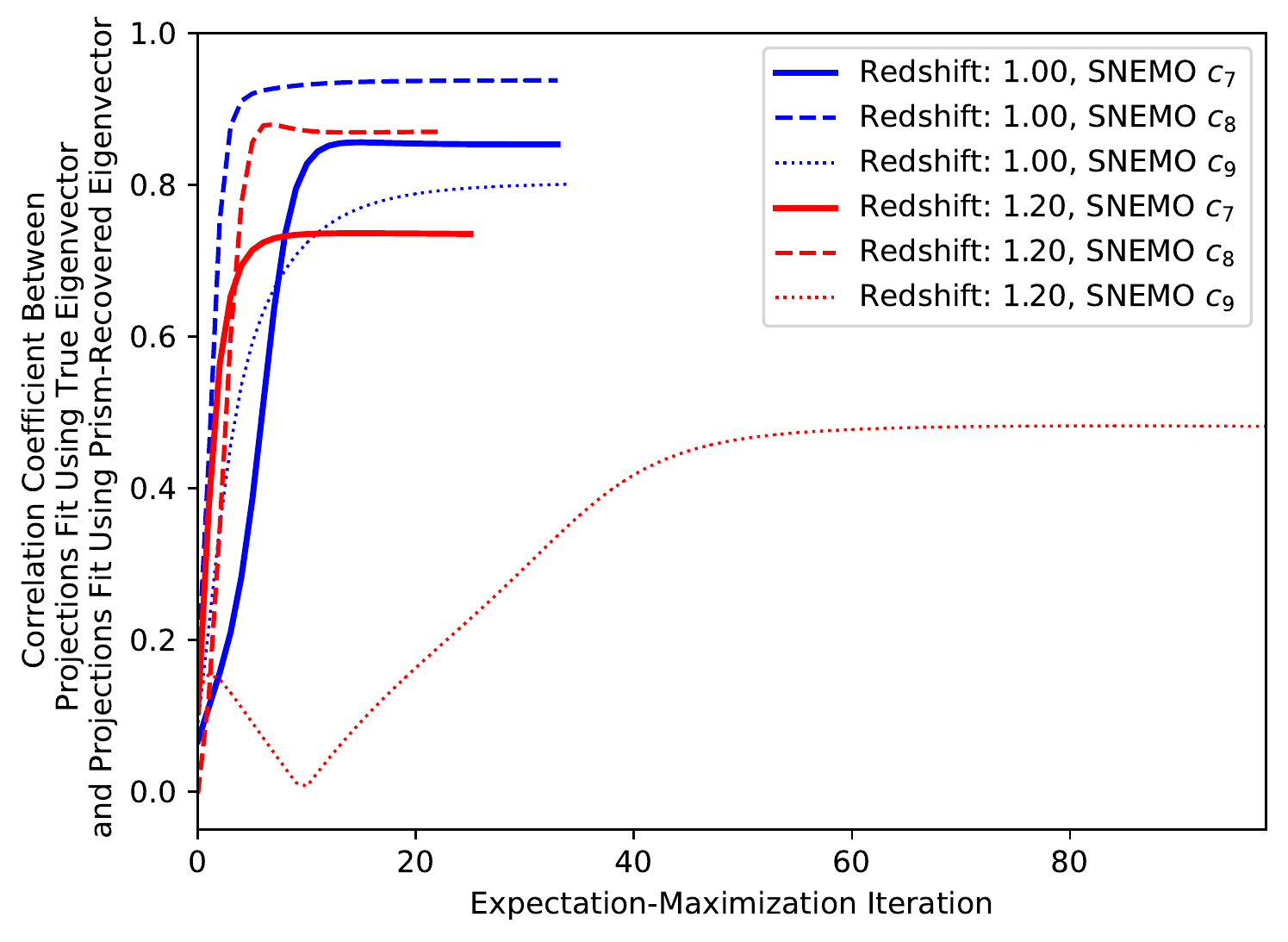}
    \caption{Using six independent analyses of 100 SNe, we investigate the convergence of an \gls{EMPCA} in recovering each of three different unknown eigenvectors (\gls{SNEMO}15's seventh, eighth, and ninth) at each of two different redshifts (1 and 1.2). We monitor the convergence by looking at the correlation coefficient between the eigenvector projections using the true eigenvector, and the eigenvector projections using the best estimate of the missing eigenvector at each iteration. We see high correlations at redshift 1 and somewhat lower correlations for simulations at redshift 1.2.}
    \label{fig:EMPCA}
\end{figure}

Figure~\ref{fig:EMPCA} shows the progression of the iterations in terms of the correlation coefficient between the projections found with the true eigenvector, and the projections found in the e-step with the estimated eigenvector. This figure shows three different eigenvectors (\gls{SNEMO}'s 7, 8, and 9) which are neither the most obvious and easiest to constrain, nor the least obvious (cf. Figure~\ref{fig:prism_subtype}). For the sample of 100 simulated SNe at redshift 1, we see rapid convergence and high correlation coefficients; for the sample of 100 simulated SNe at redshift 1.2, we see worse performance comparing the same eigenvectors.

\begin{figure}[htbp]
    \centering
    \includegraphics[width=0.65 \textwidth]{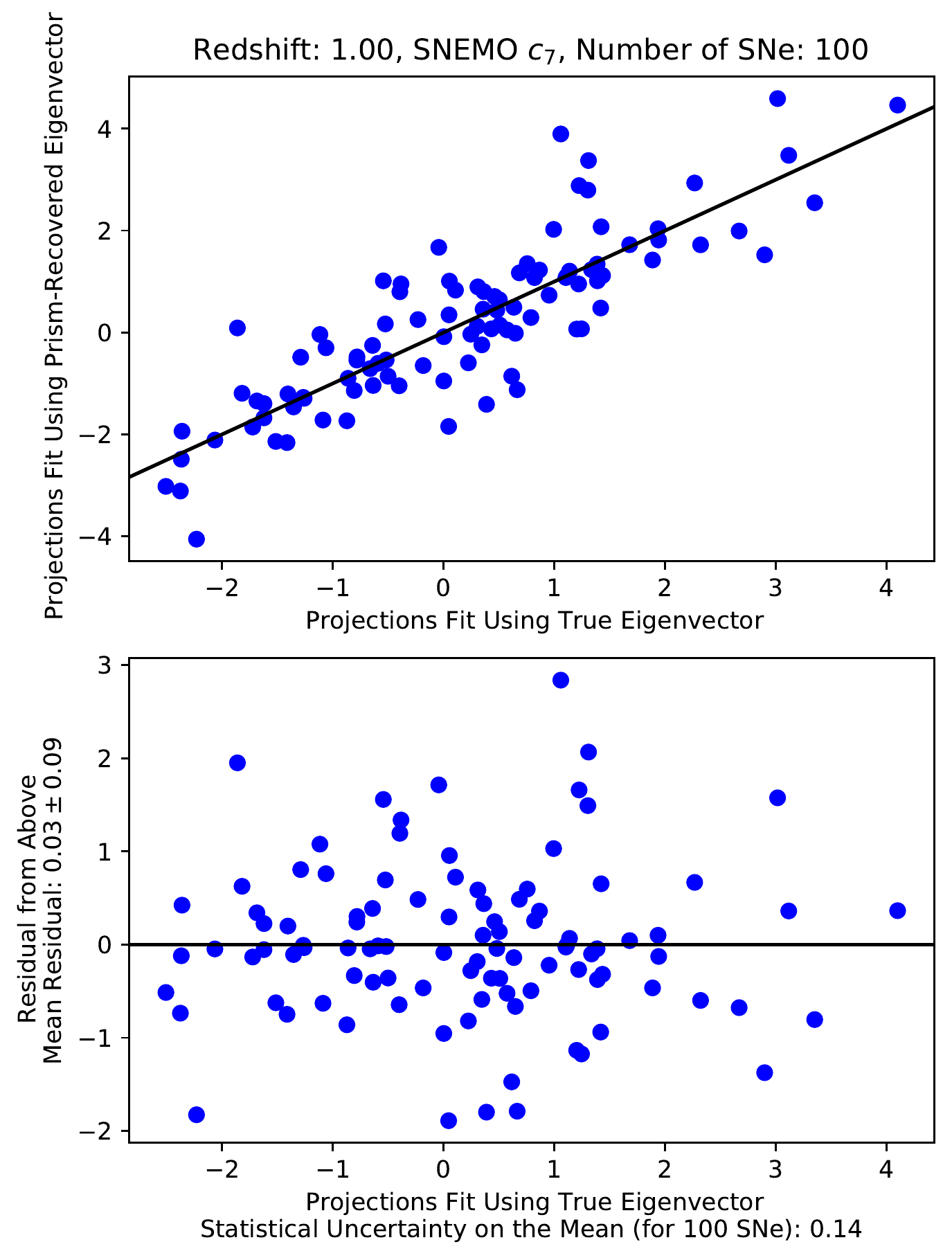}
    \caption{{\bf Top panel}: Estimated eigenvector projections for the 7th \gls{SNEMO}15 eigenvector recovered from 100 simulated prism time series at redshift 1. This is a typical case from Figure~\ref{fig:EMPCA}. The x-axis shows the projections estimated with perfect knowledge of the eigenvector; the y-axis shows the projections when the eigenvector is estimated from an \gls{EMPCA}. This panel shows reasonable agreement. Note that no noise is directly seen in this, as the same simulated time series are used for the projections shown on both axes (the noise does propagate into the recovered eigenvector). {\bf Bottom panel:} the residuals from the top panel, showing that the mean difference between the mean difference in projections constrained to somewhat better than the statistical uncertainty. This indicates that an unknown eigenvector can be recognized, and its effect on SN distances measured and mostly taken into account in a cosmology analysis.}
    \label{fig:evcorrelation}
\end{figure}

Figure~\ref{fig:evcorrelation} shows a typical recovery. The top panel shows the projections one finds with the recovered eigenvector plotted against the projections one finds with perfect eigenvector knowledge. The bottom panel shows the residuals. The scaling of the eigenvector is arbitrary, i.e., multiplying the eigenvector by two and dividing all the projections by 2 will leave the fluxes the same. For the purposes of this figure, we scale the eigenvector so that the bottom panel has zero correlation with the projections using the true eigenvector. Note also that there is no noise directly on this figure (although noise does propagate into the recovered eigenvector), as the same simulated time series is fit for both the values on the x-axis and the values on the y-axis.

As shown in Figure~\ref{fig:evcorrelation}, we obtain almost the same eigenvector projections using our recovered eigenvector as the real one. Thus, if an unknown eigenvector appeared only in the prism spectra, we would be able to find it, measure its impact on luminosity, and take it into account in the cosmology analysis. The exact requirements will depend on the eigenvector, but S/N $\sim 50$ is a reasonable threshold for high enough quality data for running this test.

\subsection{Preliminary Prism Survey Optimization} \label{sec:surveyoptimization}

With approximate rest-frame $V$-band S/N targets in hand (S/N 25 for typing, and S/N 35 for subtyping), we perform a simple optimization. Following \citet{Rubin21b}, we solve for the per-epoch exposure times as a function of redshift for both S/N targets (shown in Table~\ref{tab:prismexposure}), then parameterize the number of SNe as a function of redshift (in bins of 0.1). Each survey is represented by a non-negative vector of the relative number of SNe as a function of redshift. The \citet{Rubin21b} optimizer scales the relative numbers of SNe to absolute numbers by finding the minimum number of pointings and exposure times that give those relative numbers, then scaling that survey linearly to take a certain fixed total time. This scaling can produce fractional pointings, so for small surveys with a few pointings it is only an approximation to the optimal survey. As with the \gls{SDT} report, we combine with a 0.2\% CMB shift-parameter constraint (defined in \citealt{Efstathiou1999}) and 800 nearby SNe and assume a flat $w_0$-$w_a$ cosmology to compute a \gls{FoM}. The survey is optimized by adjusting the relative numbers of SNe as a function of redshift to produce the maximum \gls{FoM}.

Table~\ref{tab:optimumsurveys} shows our optimized surveys. They have two tiers: which we refer to as ``wide'' and ``deep.'' We vary three sets of input assumptions: the S/N needed in the prism data for a SN to be useful (either S/N~25 for typing, or S/N~35 for subtyping) the Hubble diagram RMS (0.15, 0.1, or 0.075 magnitudes), and the amount of time used for the prism survey (either 0.125~years, 0.25~years, or 0.5~years). \citet{Rose2021} considered prism times amounting to 10\%, 25\%, 50\%, and 75\% of the 0.5 year \gls{HLTDS} survey. We select the middle two options, and also consider 0.5~years not as a serious proposal, but just as a comparison to imaging-only surveys. Appendix~\ref{sec:prismvsimaging} performs a preliminary optimization that suggests $\sim 25\%$ or 0.125 years for the prism is a reasonable option, but this needs further study.

Surprisingly, the optimized surveys are very similar across all of these assumptions: \prismsurveyoptimalsummary (double this for the 0.25-year surveys and quadruple for the 0.5-year surveys). We also note that the statistical-only \gls{FoM} values can be quite high ($\sim 200$--400 for a 0.125 year survey). Depending on the level of systematic uncertainty, an initial cosmology analysis using just SNe observed with both the prism and imaging may be a useful interim step towards the full cosmology analysis that also includes SNe observed in just imaging.

\begin{deluxetable}{c|cccc}[htbp]
\tablehead{
\colhead{Redshift} & \colhead{S/N $7.50\times4\ \mathrm{Dithers}=15.00$} & \colhead{$10.61\times2=15.00$} & \colhead{$12.50\times4=25.00$} & \colhead{$17.50\times4=35.00$}
}
\startdata
0.3 & 186.45 & 121.48 & 276.85 & 367.25\\
0.4 & 180.80 & 118.65 & 268.38 & 355.95\\
0.5 & 203.40 & 132.78 & 302.28 & 403.98\\
0.6 & 245.78 & 161.03 & 372.90 & 505.68\\
0.7 & 296.62 & 197.75 & 457.65 & 638.45\\
0.8 & 347.48 & 237.30 & 553.70 & 793.83\\
0.9 & 418.10 & 288.15 & 686.48 & 1031.12\\
1.0 & 494.38 & 350.30 & 844.68 & 1322.10\\
1.1 & 570.65 & 418.10 & 1025.48 & 1680.88\\
1.2 & 661.05 & 497.20 & 1243.00 & 2118.75\\
1.3 & 768.40 & 598.90 & 1528.33 & 2706.35\\
1.4 & 898.35 & 726.03 & 1884.28 & 3412.60\\
1.5 & 1042.42 & 867.28 & 2288.25 & 4234.68\\
1.6 & 1217.58 & 1048.08 & 2799.58 & 5262.98\\
1.7 & 1460.53 & 1299.50 & 3503.00 & 6667.00\\
1.8 & 1697.83 & 1542.45 & 4186.65 & 8006.05\\
1.9 & 2017.05 & 1881.45 & 5144.33 & 9927.05\\
2.0 & 2387.12 & 2262.83 & 6217.83 & 12057.10\\
\enddata
\caption{The exposure times (in seconds) per visit necessary to reach a given rest-frame $V$ S/N over the full time series of the median SN. We show S/N 15, 25, and 35, with the shorter S/N 15 exposure times with both two dithers (S/N $10.61 \times 2$) and four dithers (S/N $7.50\times 4$) to show the impact of read noise. These times do not include our assumed slew time per pointing of \slewtime.}
\label{tab:prismexposure}
\end{deluxetable}

\begin{deluxetable}{cc|ccc}[htbp]
\tablehead{
\colhead{$V$ S/N Target} & \colhead{HD RMS (Mag)}  & \colhead{Wide Tier} & \colhead{Deep Tier} & \colhead{Statistical-Only FoM}
}
\startdata
\multicolumn{5}{c}{Surveys using 0.125 Years of Prism} \\
\hline
25 & 0.15 & 5.6 deg$^2$, 457.65 s & 1.5 deg$^2$ 2799.57 s & 228 \\ 
25 & 0.10 & 6.2 deg$^2$, 457.65 s & 1.5 deg$^2$ 2799.57 s & 383 \\ 
35 & 0.10 & 4.8 deg$^2$, 638.45 s & 0.8 deg$^2$ 5262.97 s & 292 \\ 
35 & 0.075 & 6.4 deg$^2$, 505.67 s & 0.7 deg$^2$ 5262.97 s & 402 \\ 
\hline
\multicolumn{5}{c}{Surveys using 0.25 Years of Prism} \\
\hline
25 & 0.15 & 12 deg$^2$, 457.65 s & 3.0 deg$^2$ 2799.57 s & 335 \\ 
25 & 0.10 & 13 deg$^2$, 457.65 s & 2.9 deg$^2$ 2799.57 s & 544 \\ 
35 & 0.10 & 10 deg$^2$, 638.45 s & 1.5 deg$^2$ 5262.97 s & 432 \\ 
35 & 0.075 & 10 deg$^2$, 638.45 s & 1.6 deg$^2$ 5262.97 s & 584 \\ 
\hline
\multicolumn{5}{c}{Surveys using 0.5 Years of Prism} \\
\hline
25 & 0.15 & 24 deg$^2$, 553.70 s & 4.0 deg$^2$ 4248.8 s & 462 \\ 
25 & 0.10 & 29 deg$^2$, 457.65 s & 4.8 deg$^2$ 2799.57 s & 731 \\ 
35 & 0.10 & 19 deg$^2$, 638.45 s & 3.2 deg$^2$ 5262.97 s & 602 \\ 
35 & 0.075 & 24 deg$^2$, 505.67 s & 2.8 deg$^2$ 5262.97 s & 806 \\ 
\enddata
\caption{Simple optimized survey strategies and \gls{FoM} values. Each row shows a set of assumptions ($V$ S/N targets corresponding with Table~\ref{tab:prismexposure}, and Hubble diagram RMS), the optimized strategy, and the statistical-only \gls{FoM}. Surprisingly, the optimum survey strategies look similar (\prismsurveyoptimalsummary), irrespective of assumptions. The statistical-only \gls{FoM} values of just these-prism observed SNe can be quite high, even in the 0.125-year prism survey, implying that an initial cosmology analysis using just SNe observed with the both the prism and imaging may be a useful interim step towards the full cosmology analysis including imaging-only SNe.}
\label{tab:optimumsurveys}
\end{deluxetable}

\clearpage

\subsection{Prism Parameter Optimization} \label{sec:prismoptimization}

The optimization of the prism consists of three related parameters (and their interaction with the optical design): the wavelength of the cutoff on the blue side, the cutoff wavelength on the red side, and the overall scaling of the dispersion as a function of wavelength. Widening the spectral range or increasing the dispersion reduce the contrast of faint continuum sources against the background and thus decrease the sensitivity. However, the goals outlined in Section~\ref{sec:introduction} (redshifts, classifications, and sub-classifications) benefit from having a wider spectral range and to some extent benefit from having higher dispersion. We performed a series of analyses varying the prism parameters and investigating the sub-typing performance at fixed exposure time, i.e., remaking Figure~\ref{fig:prism_subtype}. In short, we conclude that the blue cutoff should be set as blue as the prism image quality allows ($\sim 7500$\Ang), the red cutoff should be set to $\sim 18000$\Ang to minimize thermal background, and the dispersion should be \gooddispersion. (Appendix~\ref{sec:analytic} performs a simple analytic calculation that supports this range of dispersion.) Figure~\ref{fig:filters} shows the current design based on these parameters.

The optimization of the dispersion is involved enough that it is worth describing here. The concern with simply adequately resolving the SN spectral features (minimizing dispersion to increase S/N per unit time) is that biases will occur in the interpretation of the data if the \gls{LSF} is incorrect. An in-depth investigation of this bias necessarily involves 2D simulations, rather than just 1D S/N calculations and we describe these 2D simulations in detail in Appendix~\ref{sec:twoDforward}. We evaluate the bias by generating the time series with the \gls{PSF} provided by the Project, and fitting the time series assuming an incorrect \gls{PSF}, in this case scaling the \gls{PSF} by 0.95 in the dispersion direction (resulting in a \gls{LSF} error of 5\%). 

\begin{figure}[htbp]
    \centering
    \includegraphics[width = 0.95 \textwidth]{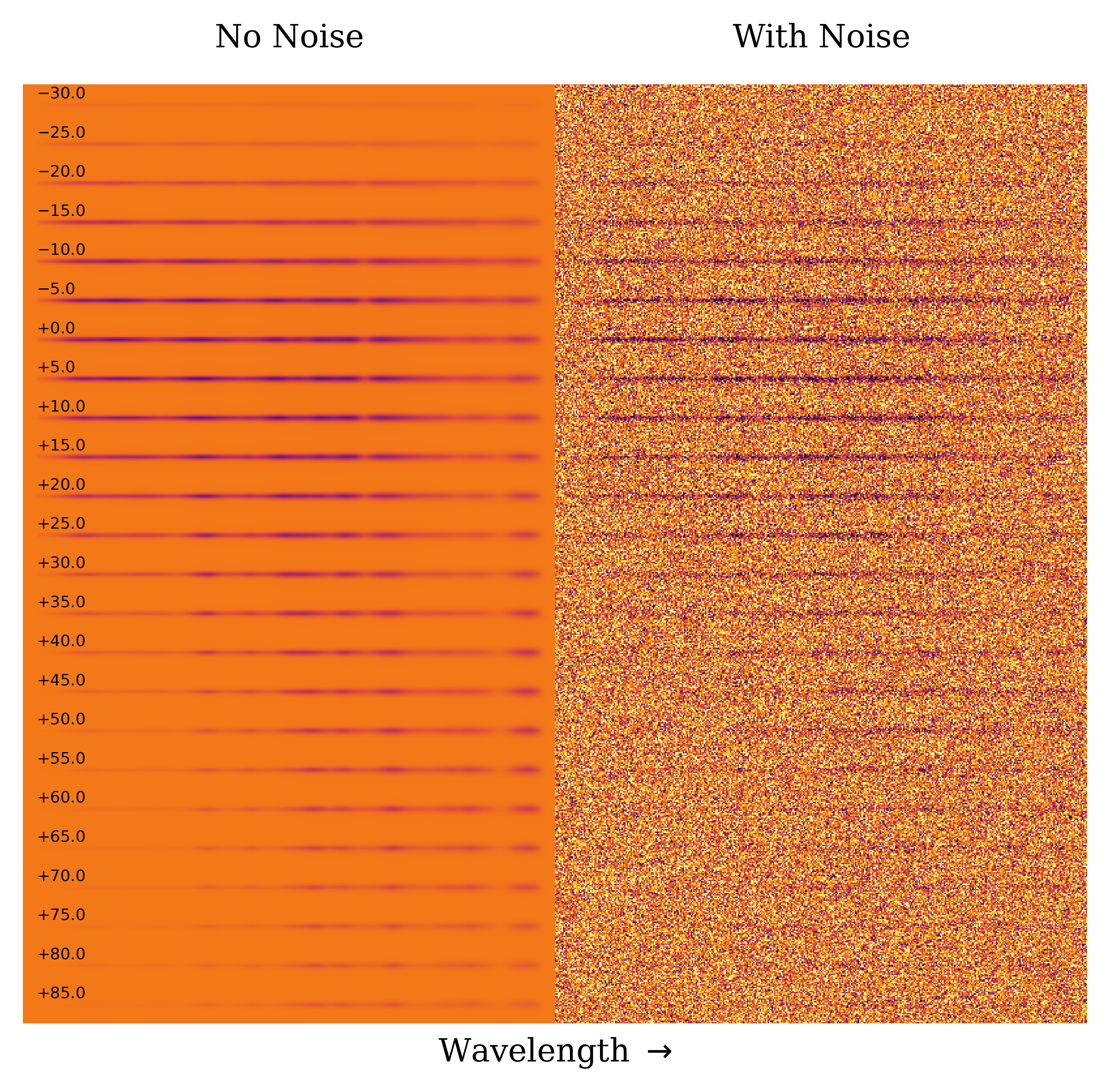}
    \caption{A simulated prism time series for a SN~Ia at redshift 1 sampled every 5 observer-frame days based on our forward-model code. The {\bf left column} shows the time series without noise added; the {\bf right column} shows noise appropriate for 1 hour visits (900-second exposures with a $2\times2$ dither pattern). (A big$\times$little$\times$little dithering strategy that spans chip gaps is likely more optimal, but $2\times2$ suffices for illustrative purposes.) The row for each date shows the four dithers interlaced. As expected from Figure~\ref{fig:SNRmax}, the S/N in any one visit is modest, but the S/N over the time series is quite reasonable.
    \label{fig:forwardmodelimage}}
\end{figure}

\begin{figure}[htbp]
    \centering
    \includegraphics[width=0.8 \textwidth]{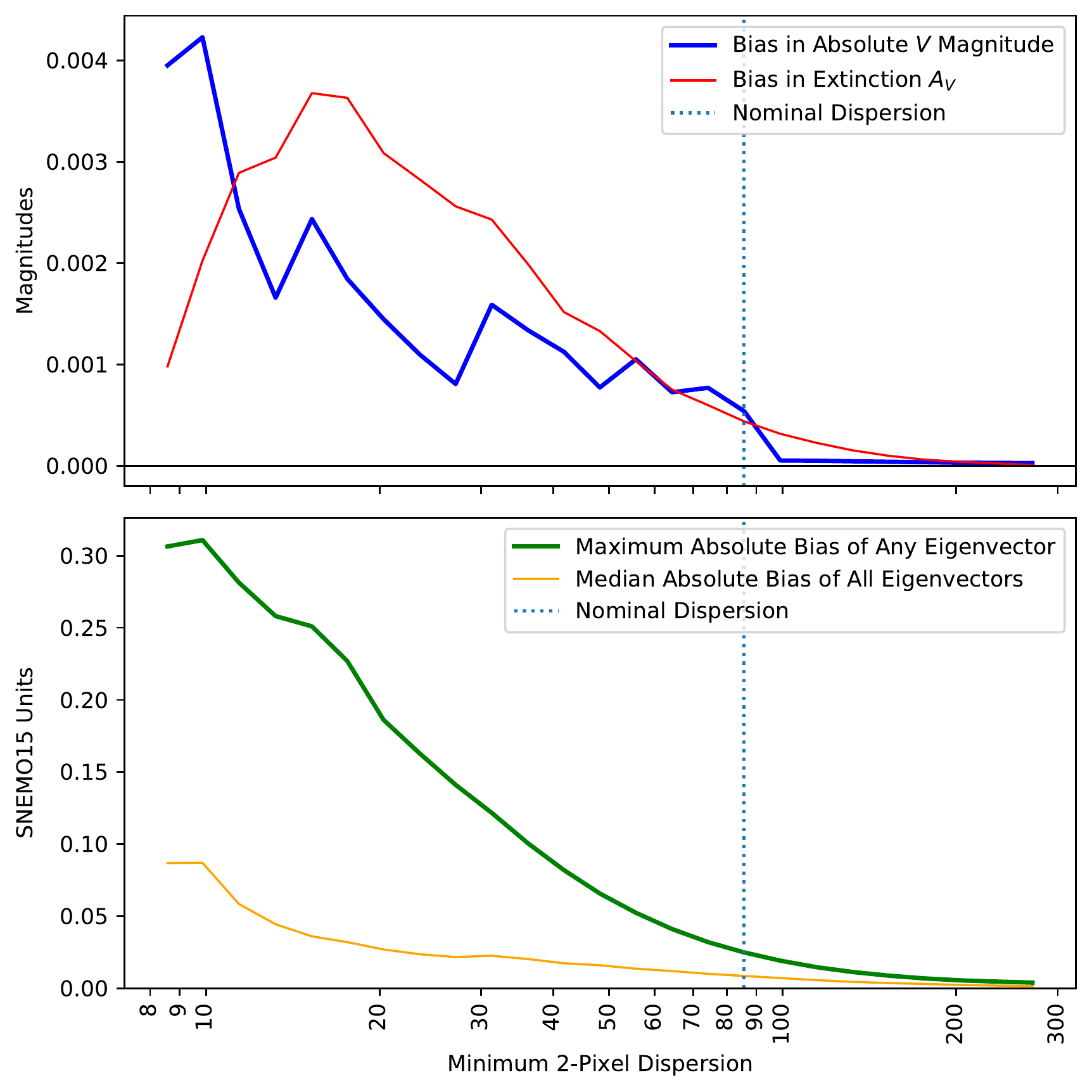}
    \caption{Biases due to an incorrect \gls{LSF}. We use our forward-model code to generate a simulated $z=1.1$ SN with the correct \gls{PSF}, then fit it using an \gls{PSF} scaled by 0.95 in the dispersion direction (for a 5\% error in the \gls{LSF}). We are only interested in measuring biases, so we do not add noise in the forward-model code, enabling us to measure the bias using only one SN time series at each dispersion value. The biases incurred drop rapidly with the dispersion, as the SN spectral features are increasingly over-resolved and thus our measurements are increasingly insensitive to the assumed \gls{LSF}. The {\bf top panel} shows biases in the recovered absolute magnitude and extinction as a function of the minimum prism dispersion. The {\bf bottom panel} shows biases in the eigenvector projections. For minimum two-pixel dispersion of \gooddispersion (a criterion met by the nominal dispersion, shown with a dotted line), the biases are $\lesssim 1$~mmag, and the eigenvector projections are recovered to a few percent of the intrinsic distribution widths ($\sim 1$). Thus with $\sim 5$\% uncertainty in the \gls{LSF}, we can plausibly average over $\sim 1000$ SNe to look for any biases without being limited by systematic uncertainties.}
    \label{fig:bias_vs_R}
\end{figure}

Figure~\ref{fig:forwardmodelimage} shows a time series simulated for this purpose at $z=1$. We use the time series without noise added (left column) to directly and precisely evaluate the bias without having to average over many thousands of noisy SNe. Figure~\ref{fig:bias_vs_R} shows the biases one obtains as a function of dispersion fitting with the incorrect \gls{PSF}. For a minimum dispersion \gooddispersion, the biases are modest.

\section{Conclusion} \label{sec:conclusion}

This work presents a series of studies investigating the uses of a low-dispersion prism in the \WFIRST mission (now baselined). Broadly speaking, many of our studies are idealized, assuming perfect host-galaxy and background subtraction (except for the impact of Poisson noise), relying on existing SN \gls{SED} models (each of which has significant limitations), assuming a good calibration, and ignoring the details of the survey geometry. Future work will introduce more realism in simulation and treatment of the prism data. But these studies do indicate that the prism can produce data with S/N and wavelength sampling that is useful for a broad range of SN investigations.

We find that using such a prism for part of the \WFIRST \gls{HLTDS} provides crucial SN data for a significant and representative sample of SNe~Ia that would be difficult to otherwise obtain. We perform a simple survey optimization, and present a toy survey that shows what performance is possible with exposure times in the range \wideexp--\deepexp seconds. We show that live-SN redshifts from such a survey extend above redshift 2, SN Ia subclassification is possible to \redshiftsubtype, and useful \gls{SED} training information is available at redshift 1--1.2. In short, we find the prism addresses many of the systematic uncertainties that are present in an imaging-only survey. Future work will also seek to continue to optimize the prism component of the \gls{HLTDS}, including the relative amount of time spent performing imaging and spectroscopy.

\clearpage

\printglossaries

\begin{acknowledgments}
This work was supported by NASA through grant NNG16PJ311I (Perlmutter \WFIRST Science Investigation Team). This work was also partially supported by the Office of Science, Office of High Energy Physics, of the U.S. Department of Energy, under contract no. DE-AC02-05CH11231. L.G. acknowledges financial support from the Spanish Ministerio de Ciencia e Innovaci\'on (MCIN), the Agencia Estatal de Investigaci\'on (AEI) 10.13039/501100011033, and the European Social Fund (ESF) ``Investing in your future'' under the 2019 Ram\'on y Cajal program RYC2019-027683-I and the PID2020-115253GA-I00 HOSTFLOWS project, from Centro Superior de Investigaciones Cient\'ificas (CSIC) under the PIE project 20215AT016, and the program Unidad de Excelencia Mar\'ia de Maeztu CEX2020-001058-M.

\end{acknowledgments}

\software{
Astropy \citep{Astropy},
Concorde \citep{Concorde},
Matplotlib \citep{matplotlib}, 
Numpy \citep{numpy}, 
SciPy \citep{scipy},
SNCosmo \citep{sncosmo},
}

\appendix

\section{Prism Simulations} \label{sec:prismsimulations}

\subsection{S/N Calculations with 1D Simulations} \label{sec:OneDModel}

For the 1D simulations, we take (as a function of wavelength) the prism PSFs from the Project (convolved with a square 0\farcs11 pixel),\footnote{Usually we use the Prism\_3Element\_PSF\_SCA08 PSFs, as these are roughly at the median image quality over the field of view.} the effective area,\footnote{Roman\_effarea\_20201130.xlsx} and the dispersion.\footnote{GRISM\_PRISM\_Dispersion\_190510.xlsx} We convert the dispersion to discrete wavelengths ($\lambda$) and the wavelength range spanned by each pixel ($\Delta \lambda$). We convert the source flux to photoelectrons per wavelength pixel with
\begin{equation}
    \mathrm{Photoelectrons\ Per\ Wavelength\ Pixel} = \mathrm{Source\ Flux\ (in\ ergs/cm}^2\mathrm{/s/\Ang)} \times \Delta \lambda \times t_{\mathrm{exp}} \times \frac{\lambda}{hc} \times \mathrm{Effective\ Area}\ ,
\end{equation}
where $t_{\mathrm{exp}}$ is the exposure time, and $hc/\lambda$ is the energy per photon. Then we assemble a simulated spectrum. We linearly interpolate the PSFs to each wavelength, and conservatively throw out the PSF more than $\pm$ 0\farcs25 parallel to the prism trace (in principle, a forward-model code with excellent line-spread function knowledge may recover some of this lost S/N). Then we collapse the remaining PSF parallel to the trace, and place the photoelectrons from the source on each set of pixels in wavelength.

The noise comes mostly from the Poisson noise of zodiacal light, but we also include the SN, its host galaxy, thermal, read noise, and dark current. Integrating the zodiacal model of \citet{aldering02} (appropriate for an ecliptic latitude of 75$^{\circ}$) over the effective area of the prism, we find 1.274 electrons/pixel/s of background. As described below, we increase this by 5\% for host-galaxy background (described below) and add 0.003 electrons/pixel/s for detector dark current and instrumental backgrounds. We use the thermal model of \citet{Rubin21b}, giving 0.012 electrons/pixel/s for a total background of 1.353 electrons/pixel/s. Read noise only contributes to short exposures, but we use 15 electrons per 2.825~second readout, and a floor of 5 electrons.

For the noise from contamination by other objects, we consider the host galaxy, as it is (almost always) the nearest object and underneath the SN. Typical host-galaxy surface brightnesses are close to zodiacal \citep{Hounsell2018}, but the brightest part of the host galaxy will generally be only $\sim$~9 pixels in size ($\sim 1\arcsec$ or $\sim 8$~kpc) while the SN spectrum is $\sim 200$~pixels long. Thus the galaxy will contribute typically 5\% of the zodiacal light. However, the galaxy spectrum will not be flat (unlike the zodiacal, which will be almost the same in every pixel). For the moment, we increase our zodiacal by 5\% and save more detailed simulations for future work \citep{Astraatmadja22}.

The extractions are simple optimal extractions where the PSF is assumed known, and the number of sky pixels is much larger than the number of SN-illuminated pixels, so we can ignore the uncertainty in the level of the zodiacal background (we do not ignore the sky Poisson noise). The estimated flux is:
\begin{equation}
    \widehat{\mathrm{flux}} = \frac{
    \sum_i \frac{1}{\sigma_i^2} P_i f_i
    }{
    \sum_i \frac{1}{\sigma_i^2} P_i^2
    }\;,
\end{equation}
and the uncertainty is
\begin{equation}
    \widehat{\sigma_\mathrm{flux}} = \frac{1}{\sqrt{
    \sum_i \frac{1}{\sigma_i^2} P_i^2
    }}\;.
\end{equation}
All sums are evaluated over pixels $i$ transverse to the dispersion, and the $P$ is the PSF collapsed parallel to the trace (so it is only one-dimensional).

\begin{figure}[htbp]
    \centering
    \includegraphics[width=0.6 \textwidth]{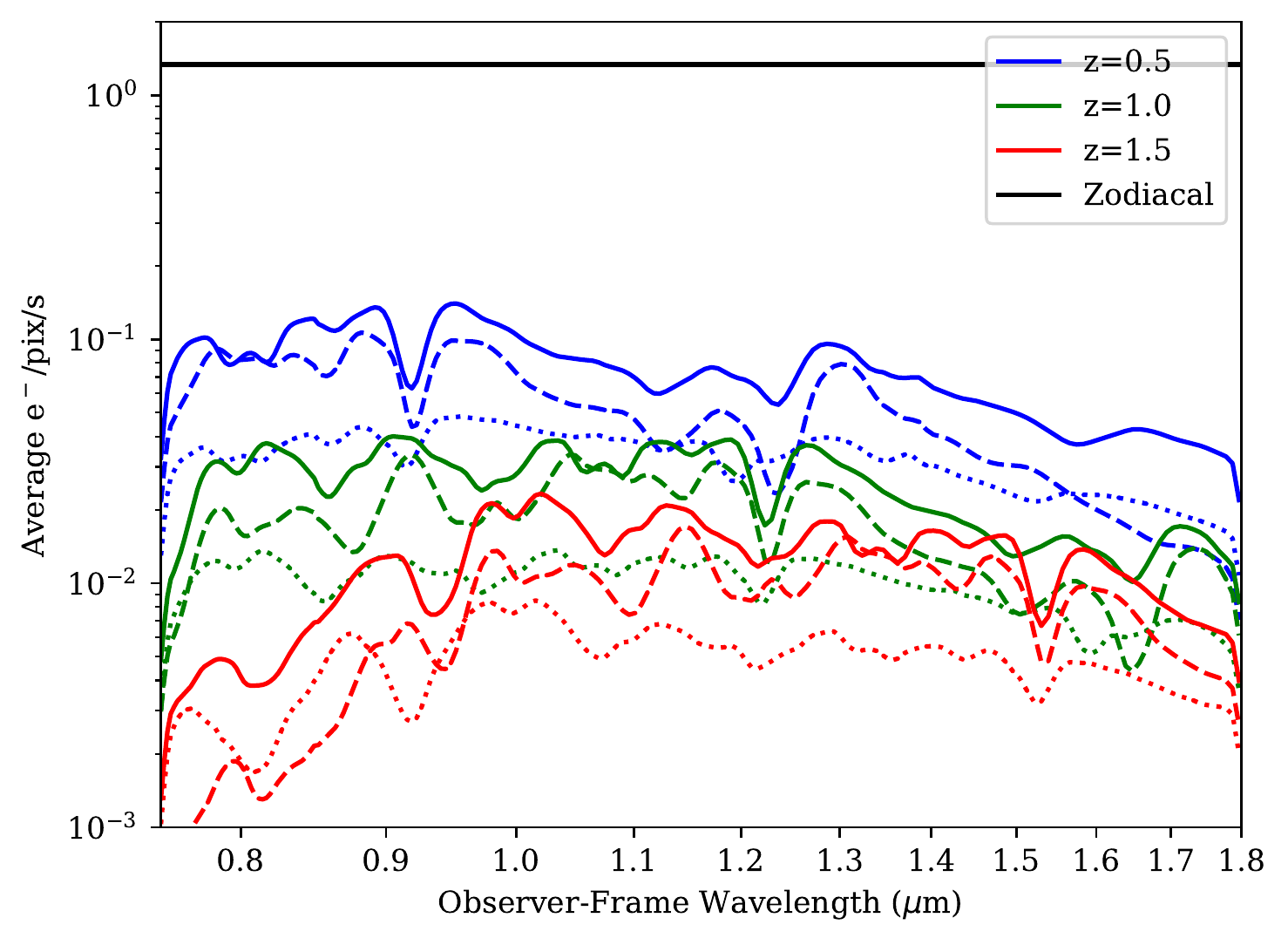}
    \caption{e$^-$/pix/second for SNe at three redshifts, illustrating contrast ratio against the sky background. The dotted line is at $-10$ days in the rest frame, the solid line is at maximum light, and the dashed line is at +10 days. High-redshift SNe can be 1\% or less of the sky background, so careful sky subtraction, flat fielding, and dithering may be necessary to achieve the full performance of the prism.}
    \label{fig:eminuspersec}
\end{figure}

Figure~\ref{fig:eminuspersec} illustrates the PSF-weighted e$^-$/s/pixel for both the zodiacal background and SNe Ia at a range of redshifts and phases. We see that the SN light varies from a fraction of 1\% to 10\% of the background. As the S/N per dither per pixel for high-redshift SNe will be $\sim$ a few, the sky subtraction/flat fielding will have to be known to better than 0.1\% per pixel in order for the flat field to not contribute noise. Uncertainties that are correlated over the length of the spectra (e.g., from scattered light) will have to be less than $\sim$ 0.01\% over most focal-plane positions. The large number of dithers and rotations inherent in the observing strategy will likely help to suppress such uncertainties, but this should be quantified in the future.

\subsection{2D Forward Model} \label{sec:twoDforward}

To address the sensitivity vs. bandwidth/dispersion tradeoff in detail, we wrote a forward-model code for simulating and extracting prism spectra. This analysis is similar in philosophy to \citet{Bolton2010}, \citet{shukla17}, or \citet{Ryan2018}. We leave fully 3D simulations with realistic galaxy backgrounds (e.g., \citealt{Rubin2021}) to future work \citep{Astraatmadja22}. Here, we perform similar simulations as in Section~\ref{sec:OneDModel}, but make true 2D images (not collapsing the PSF parallel to the trace) and choose a $5\times$ smaller wavelength bin $\Delta \lambda$ to preserve sub-pixel information. Some of this information can be recovered by dithering parallel to the trace, and we simulate a four-point $2\times 2$ dither pattern.

For extracting the spectra, we directly fit the \gls{SED} model to the simulated data (which is optimal, in principle). Extracting 1D spectra is also possible in principle, but would require computing the effective resolution of the instrument \citep{Bolton2010}.

\section{Analytic Optimization of a Background-Limited Spectrograph for Measuring Gaussian Spectral Features} \label{sec:analytic}

\newcommand{\sigbeta}{\sigma_{\beta}\xspace}
\newcommand{\sigLSF}{\sigma_{\mathrm{LSF}}\xspace}
\newcommand{\sigbetao}{\sigma_{\beta 0}\xspace}

For a sanity check on our recommended prism parameters, this section derives the optimum dispersion for a background-limited spectrograph. We assume that the line profile to be measured is a well-isolated Gaussian, as is the \gls{LSF} of the instrument (including convolution with the pixel). We also assume that the dither spacing and/or pixel spacing oversamples the spectrum. Without loss of generality, we assume the per-wavelength-pixel uncertainties are 1 (and that the spectrograph is background-limited, so these uncertainties do not change over the spectral feature). We can start by writing the $\chi^2$ as:
\begin{equation} \label{eq:simplechi2}
    \chi^2 = \sum_i \left[ \frac{\mathrm{observations} - \mathrm{model}}{\mathrm{uncertainty} = 1} \right]^2 \;.
\end{equation}
The standard deviation in pixels of the spectral feature is the convolution of the \gls{LSF} ($\sigLSF$) and the intrinsic width of the spectral feature divided by the speed of light ($\sigbetao$) times the one-pixel dispersion of the instrument ($2R$). For example, a spectral feature of standard deviation 3,000 km/s ($\sigbetao = 0.01$) will have a standard deviation of at least two pixels as dispersed by a spectrograph with a two-pixel dispersion $R = 100$ ($2R = 200$). The mean location of the spectral feature in pixels will also scale as the mean velocity divided by the speed of light ($\beta_0$) times $2R$. Finally, we make a distinction between the parameters for amplitude ($A$), mean location ($\beta$) and width ($\sigbeta$) and the true values ($A_0$, $\beta_0$, and $\sigbetao$, respectively). With all this in hand, we can write the $\chi^2$ of Equation~\ref{eq:simplechi2} as
\begin{equation}
  \chi^2 = \sum_i \left[ \frac{A_0}{\sqrt{2 \pi [(2 R \sigbetao)^2 + \sigLSF^2]}} \exp{ \left( \frac{-(x_i - 2 R \beta_0)^2}{2 [(2 R \sigbetao)^2 + \sigLSF^2] } \right)} - 
  \frac{A}{\sqrt{2 \pi [(2 R \sigbeta)^2 + \sigLSF^2]}} \exp{ \left( \frac{-(x_i - 2 R \beta)^2}{2 [(2 R \sigbeta)^2 + \sigLSF^2] } \right)}
  \right]^2 \;.
\end{equation}
As we are assuming the spectrum is well sampled, we can (up to an overall multiplicative factor) transform this sum over pixels into an integral over all $x$:
\begin{equation}
  \chi^2 = \int  \left[ \frac{A_0}{\sqrt{2 \pi [(2 R \sigbetao)^2 + \sigLSF^2]}} \exp{ \left( \frac{-(x - 2 R \beta_0)^2}{2 [(2 R \sigbetao)^2 + \sigLSF^2] } \right)} - 
  \frac{A}{\sqrt{2 \pi [(2 R \sigbeta)^2 + \sigLSF^2]}} \exp{ \left( \frac{-(x - 2 R \beta)^2}{2 [(2 R \sigbeta)^2 + \sigLSF^2] } \right)}
  \right]^2 dx \ .
\end{equation}
This integral results in a $\chi^2$ of
\begin{equation} \label{eq:chiintegrated}
    \chi^2 = \frac{1}{2 \sqrt{\pi}} \left[ \frac{A^2}{\sqrt{4 R^2 \sigbeta^2+\sigLSF^2}}-\frac{2 A A_0 e^{-\frac{R^2 (\beta - \beta_0)^2}{2 R^2 \left(\sigbeta ^2+\sigbetao^2\right)+\sigLSF^2}}}{\sqrt{2 R^2 \left(\sigbeta ^2+\sigbetao^2\right)+\sigLSF^2}}+\frac{A_0^2}{\sqrt{4 R^2 \sigbetao^2+\sigLSF^2}} \right] \ .
\end{equation}
We compute the inverse covariance matrix of the parameters $A$, $\beta$, and $\sigbeta$ by taking one half the Hessian matrix of Equation~\ref{eq:chiintegrated}, then evaluating this matrix at the true values $\{A,\ \beta,\ \sigbeta\} \rightarrow \{A_0,\ \beta_0,\  \sigbetao\}$. We invert this inverse covariance matrix to obtain the covariance matrix and take the diagonals, giving the parameter variances:
\begin{equation}
\{\sigma^2_A,\ \sigma^2_{\beta},\ \sigma^2_{\sigma_{\beta}} \} = \{ 3\sqrt{\pi (\sigLSF^2 + 4 R^2 \sigbetao^2)},\ \frac{\sqrt{\pi} (\sigLSF^2 + 4 R^2 \sigbetao^2)^{3/2}}{A_0^2 R^2},\ \frac{\sqrt{\pi} (\sigLSF^2 + 4 R^2 \sigbetao^2)^{5/2} }{4 A_0^2 R^4 \sigbetao^2} \}
\end{equation}
We take the derivatives of the variances with respect to $R$ to find the optimum values. For $A$, we find an optimum value of zero, i.e., minimizing the dispersion is optimal (as long as the spectrum is oversampled, which we have been assuming). For $\beta$, we find an optimum dispersion $R^*$ of 
\begin{equation}
    R^* = \frac{\sigLSF}{\sqrt{2} \sigbetao } \ ,
\end{equation}
e.g., 53 for $\sigbetao c$ = 4,000 km/s and $\sigLSF = 1$ (roughly correct for the prism \gls{LSF}, if one takes the FWHM shown in the right panel of  Figure~\ref{fig:filters} and divides by $\sqrt{8 \log{2}}$ to convert from a FWHM to $\sigma$). For $\sigbeta$, we find an optimum of 
\begin{equation}
    R^* = \frac{\sigLSF}{\sigbetao } \ ,
\end{equation}
e.g., 75 for $\sigbetao c$ = 4,000 km/s. As with every $R$ used by the \WFIRST Project, our dispersions $R$ are two-pixel dispersions, i.e., one half the one-pixel dispersions in, e.g., \citet{Foley2013} or \citet{shukla17}. To summarize, analytic optimization produces a reasonable sanity check on the prism dispersion.

\section{Information Content of Imaging and Prism Observations at Fixed Total Time} \label{sec:prismvsimaging}

This section shows a simple trade between the imaging and the prism at fixed total exposure time. We evaluate SN uncertainties using the \gls{SUGAR} model, as it has three intrinsic variability parameters ($q_1$, $q_2$, and $q_3$, plus color), so it likely has higher fidelity than \gls{SALT2} (one intrinsic variability parameter plus color), but does not have so many parameters that constraining them with light curves is hopeless.

As the prism observations are Poisson-limited and much longer than the slew times, they can be divided among all SNe or concentrated on a subset of SNe without much change in the total signal to noise across all SNe. For example, one could observe one quarter of the SNe at a certain S/N per object, but this has similar total S/N to observing the full sample at half that S/N per object. Thus for this simplistic analysis, we assume that all SNe are observed with both the imaging and the prism.

Following \citet{Rose2021}, our 100\% imaging survey consists of 300~s exposures in $F087$ ($F087$ was not included in the \citealt{Rose2021} deep tier, but is added here to strengthen the imaging constraints), $F106$, $F129$, and $F158$, with a 900~s $F184$ exposure (we follow the image simulation procedure of \citealt{Rubin21b} to make the light curves). Including slew times of \slewtime, this gives \imagingonlyseconds per pointing. This survey is then scaled to other values of prism fraction, and the \gls{SUGAR} uncertainties evaluated. We consider all values of prism fraction from 0\% prism (100\% imaging) to 100\% prism (0\% imaging), which devotes \prismonlyseconds to the prism (taking the slew time into account). Our simulations do not assume any host-galaxy-subtraction noise, and so imply that 3D forward modeling is used for the prism data (so that all the observations without SN light can be combined into a deep reference or ``template'' image cube).

\begin{figure}[htbp]
    \centering
    \includegraphics[width=\textwidth]{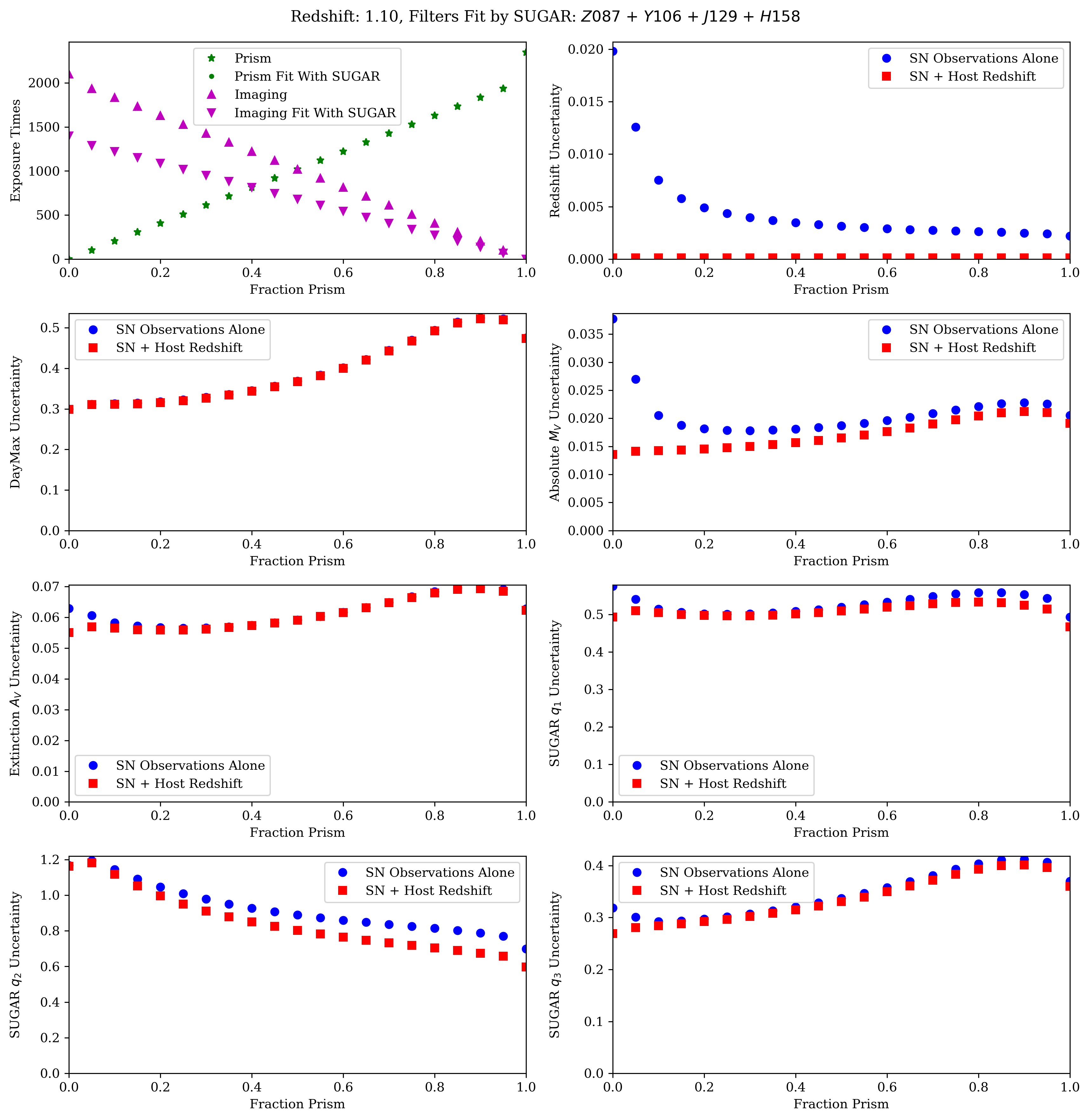}
    \caption{Evaluating SN uncertainties from SUGAR when trading imaging and prism observations at fixed total time. The {\bf top left panel} shows the imaging times for both the prism and the imaging as a function of the fraction of time spent on the prism observations. These exposure times deviate from straight lines because of overheads. The panel shows the total exposure times that are simulated (green stars for the prism) and upward-pointing magenta triangles for the imaging. Due to the limited rest-frame wavelength coverage of SUGAR, not all of the wavelengths or filters are always used in the SUGAR fitting. The total times that are actually fit are shown with green dots for the prism (overlapping the green stars at this redshift), and downward-pointing magenta triangles for the imaging. The remaining plots show the \gls{SUGAR} uncertainties, derived from a simultaneous fit of the imaging and prism time series. Both a SN-only analysis (blue circles) and a SN + host-galaxy redshift analysis (red squares) are shown. Generally speaking, going from imaging-only to $\sim 25\%$ prism time decreases or does not increase the uncertainties, with \gls{SUGAR} $q_2$ especially benefiting from having increased prism time.}
    \label{fig:ImagingPrismOneOne}
\end{figure}

\begin{figure}[htbp]
    \centering
    \includegraphics[width=\textwidth]{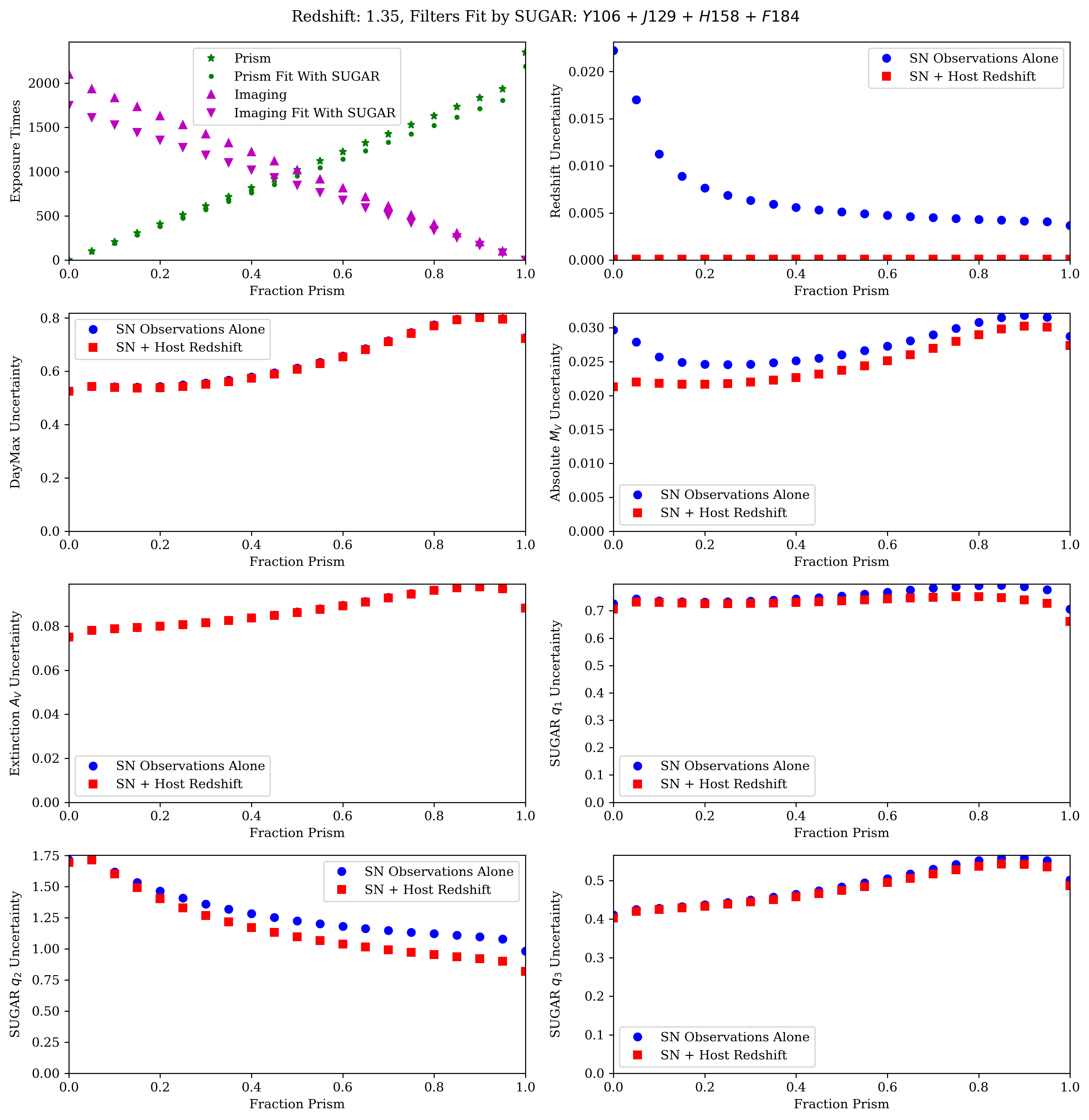}
    \caption{Similar to Figure~\ref{fig:ImagingPrismOneOne}, but redshift 1.35.}
    \label{fig:ImagingPrismOneThreeFive}
\end{figure}

\begin{figure}[htbp]
    \centering
    \includegraphics[width=\textwidth]{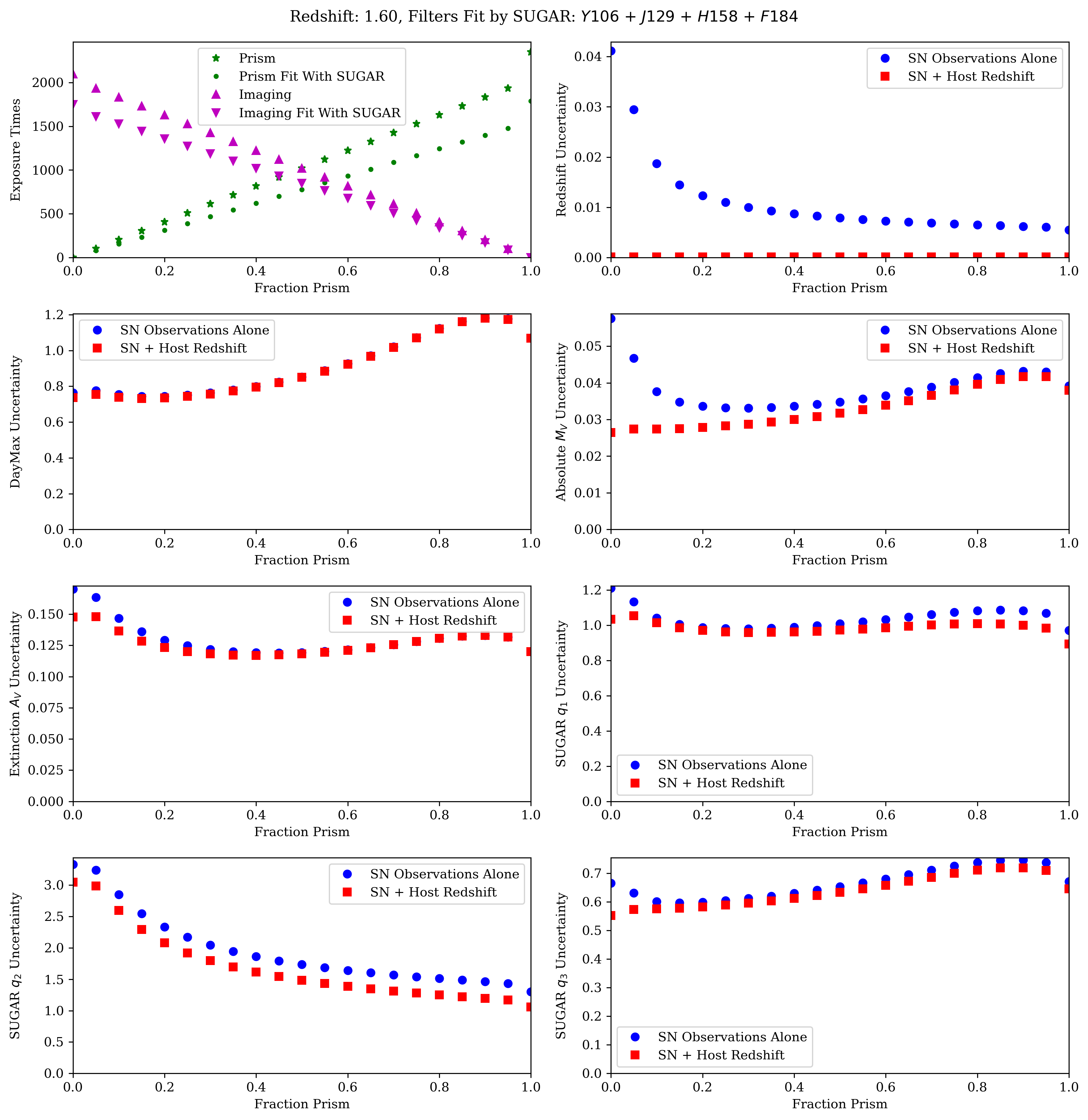}
    \caption{Similar to Figure~\ref{fig:ImagingPrismOneOne}, but redshift 1.60.}
    \label{fig:ImagingPrismOneSix}
\end{figure}

\begin{figure}[htbp]
    \centering
    \includegraphics[width=\textwidth]{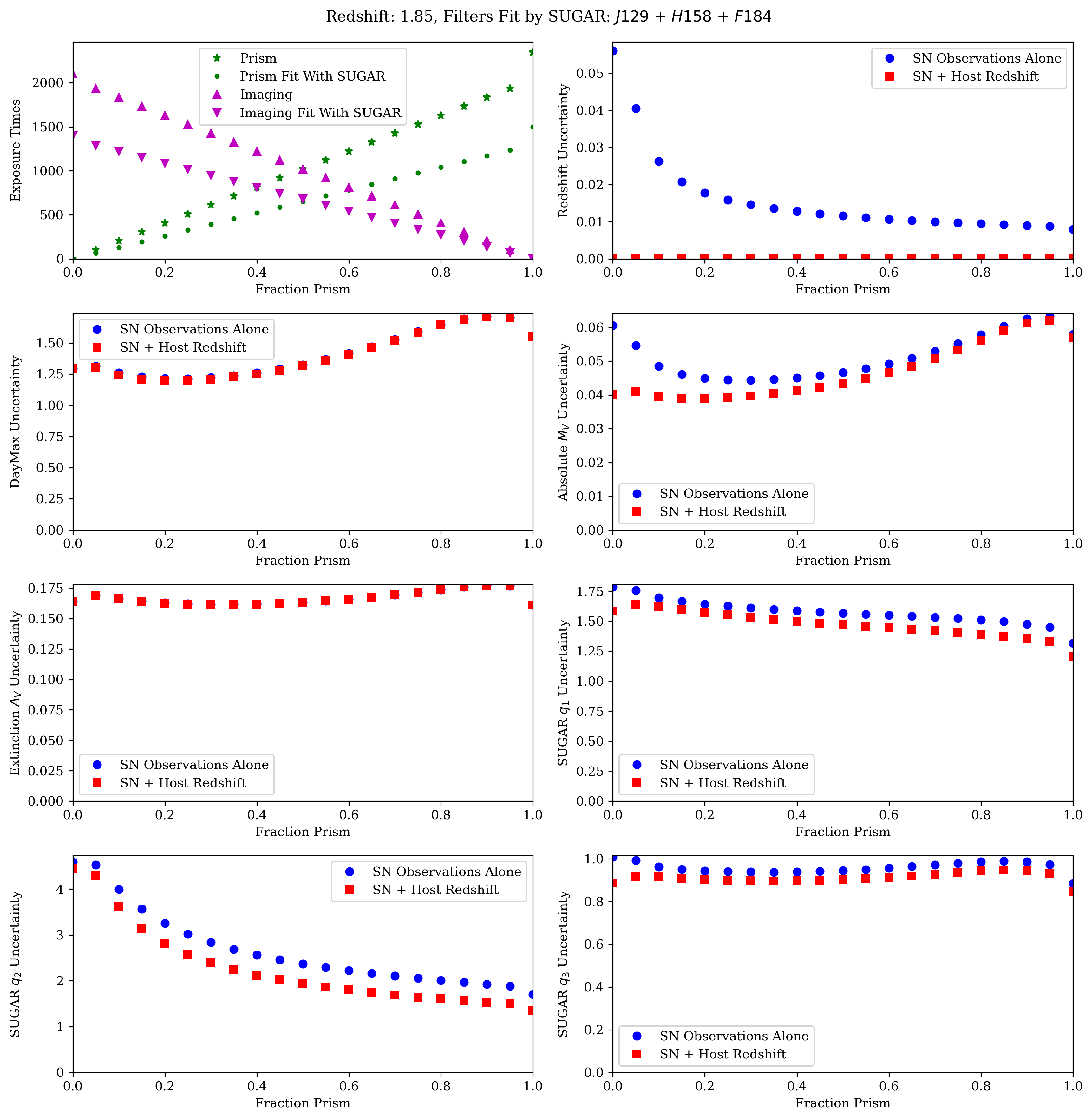}
    \caption{Similar to Figure~\ref{fig:ImagingPrismOneOne}, but redshift 1.85.}
    \label{fig:ImagingPrismOneEightFive}
\end{figure}

Figures~\ref{fig:ImagingPrismOneOne}, \ref{fig:ImagingPrismOneThreeFive}, \ref{fig:ImagingPrismOneSix}, and \ref{fig:ImagingPrismOneEightFive} show the \gls{SUGAR} uncertainties as a function of prism fraction at four different redshifts. These figures show the uncertainties with and without assuming a known host-galaxy redshift. In general, going from zero prism to $\sim 25$\% decreases or does not increase the uncertainties. The \gls{SUGAR} $q_2$ parameter is especially sensitive to the prism fraction, rapidly decreasing as time is moved from the imaging to the prism. Unfortunately, no published standardization coefficients for \gls{SUGAR} exist, so we cannot turn these uncertainties into distance-modulus uncertainties. (The \citet{Boone2021a} parameterization also has only three parameters describing the range of SN~Ia behavior, and it does provide a translation from these parameters to relative distance modulus. This parameterization is currently being developed into a full time-series model \citep{Dixon2022} and this will allow a more complete optimization study.) These results are suggestive that the prism can be a value add to the \gls{HLTDS} at fixed total exposure time, but future work will be necessary to further optimize the prism fraction.

\end{document}